\documentclass[preprint,review,12pt,a4paper,3p]{elsarticle}
\usepackage[]{geometry}
\usepackage[scriptsize,tight]{subfigure}
\usepackage{graphicx}
\usepackage{multirow}
\usepackage{dcolumn}
\usepackage{bm}
\usepackage{color,soul}
\usepackage[mathlines]{lineno}
\usepackage{hyperref}

\biboptions{longnamesfirst,sort&compress,semicolon}

\makeatletter
\def\ps@pprintTitle{%
	\let\@oddhead\@empty
	\let\@evenhead\@empty
	\let\@oddfoot\@empty
	\let\@evenfoot\@oddfoot
}
\makeatother

\begin{document}

\begin{frontmatter}
	\author{C. Aguiar}

	\author{I.~Camps}
	\ead{icamps@unifal-mg.edu.br}
	\cortext[coric]{Corresponding author}

	\address{Laborat\'orio de Modelagem Computacional - \emph{La}Model,
		Instituto de Ci\^{e}ncias Exatas - ICEx. Universidade Federal de Alfenas -
		UNIFAL-MG, Alfenas, Minas Gerais, Brazil}

	\title{Are boron-nitride nanobelts capable to capture greenhouse gases?}

	\begin{abstract}
		Why is the question in the title pertinent? Toxic gases, which are harmful to human health and the environment, have been released in greater quantities as a result of industrial development. These gases require capture, immobilization, and measurement. Consequently, the present study investigates the interactions between the boron-nitride nanobelt and the Möbius-type boron-nitride nanobelt and nine greenhouse gases, namely ammonia, carbon dioxide, carbon monoxide, hydrogen sulfide, methane, methanol, nitric dioxide, nitric oxide,
		and phosgene. The adsorption energies calculated for the structures with optimized geometry are all negative, suggesting that all gases are adsorbed favorably in both nanobelts. Furthermore, we discovered that the recovery time of the sensors ranges from 2 hours to a few nanoseconds and that the nanobelts exhibit distinct responses to each gas. According to electronic and topological investigations, covalent bonds were formed exclusively by nitric oxide; the remaining gases formed noncovalent bonds. Molecular dynamics ultimately demonstrate that the interaction between a single gas molecule and the nanobelt remains consistent across the vast majority of gases, while the
		interaction between 500~gas molecules and the nanobelts functions as an attraction, despite the impact of volumetric effects characteristic of high-volume gases on the interaction. For the completion of each calculation, semi-empirical tight-binding methods were implemented using the xTB software. The results of our study generated a favorable response to the question asked in the title.
	\end{abstract}

\end{frontmatter}

\section{Introduction}
\label{Sec:Intro}
\newcommand{\sizeA}{3.0cm}
The escalation of industrialization, the improper disposal of waste and byproducts, and the combustion of fossil fuels have all contributed to an increase in environmental
pollution~\cite{Lee-Sens.ActuatorsBChem.-255-1788-2018,Manisalidis-Front.PublicHealth-8--2020,
Nool-SSM-PopulationHealth-15-100879-2021}. Humans and the environment can be severely harmed by the emission of toxic gases, including phosgene, hydrogen sulfide, ammonia, nitric oxide, methanol, methane, carbon monoxide, and carbon dioxide~\cite{Lee-Sens.ActuatorsBChem.-255-1788-2018}.

Methanol (CH\textsubscript{3}OH) is a hazardous alcohol that is frequently encountered in industrial and domestic by-products. Prolonged exposure to it can lead to severe illness and, in severe cases, to death~\cite{Holt-Intern.Med.J.-48-335-2018}. On the contrary, methane (CH\textsubscript{4}) has the potential to cause harm due to its combustible nature when inhaled in excess~\cite{Jo-Tuberc.Respir.Dis.-74-120-2013}. Carbon monoxide (CO), an odorless, colorless, and tasteless gas, is produced when carbonaceous materials are burned incompletely~\cite{Buboltz2023,Otterness-Emerg.Med.Pract.-20-1-2018}. Victims often lose consciousness before they become aware of the seriousness of their poisoning. Increased concentrations of carbon dioxide (CO\textsubscript{2}) have the potential to negatively affect human health by displacing oxygen, resulting in toxicity through asphyxiation and acidosis. Subsequently, this can cause arrhythmias and tissue damage~\cite{Schrier-Front.Toxicol.-4--2022}. Phosgene (COCl\textsubscript{2}) exhibits a high degree of reactivity with functional groups present in the epithelium of the respiratory system, resulting in cellular deterioration~\cite{Rendell-Toxicol.Lett.-290-145-2018,Pauluhn-Toxicology-450-152682-2021}.

Hydrogen sulfide (H\textsubscript{2}S) is an odorless, highly combustible gas that induces poisoning by inhalation; exposure to high concentrations results in fatal toxicity~\cite{Ng-J.Med.Toxicol.-15-287-2019}. Ammonia (NH\textsubscript{3}) is an exothermic gas that induces necrosis in body tissues through an exothermic reaction~\cite{Pangeni-AnnalsofMedicine&Surgery-82-104741-2022}. It is a highly toxic and irritating gas that causes severe tissue damage. Significant exposure to nitrogen oxide gases (NO and NO\textsubscript{2}) can cause symptoms such as coughing, shortness of breath, and possibly even acute respiratory distress syndrome~\cite{Amaducci2023}. Additionally, NO\textsubscript{2} is
capable of photochemically reacting with other pollutants to generate acid rain or ozone, thereby exacerbating its detrimental environmental
impacts~\cite{Lee-Sens.ActuatorsBChem.-255-1788-2018,
Verma-ACSSensors-8-3320-2023}].

Due to the harmful effects of these gases, it is necessary to monitor
environmental pollution to ensure levels that do not harm living
organisms~\cite{Nazemi-Sensors-19-1285-2019,
Abooali-J.Comput.Electron.-19-1373-2020,
Ahmed-R.Soc.OpenSci.-9-220778-2022}. Researchers have examined a variety of gas detection systems and have determined that two-dimensional (2D) materials with substantial surface area, adsorption
specificity~\cite{Calvaresi-J.Mater.Chem.A-2-12123-2014a,Cezar-arXiv-2023}, chemical stability~\cite{Cezar-arXiv-2023,Chang-ACSNano-4-5095-2010}, and robust electrochemical performance~\cite{Holt-Science-312-1034-2006,Cezar-arXiv-2023} may have substantial application
potential. As an illustration, phosphorene has demonstrated efficacy in the detection of gases, including NH\textsubscript{3},
SO\textsubscript{2}, NO, and NO\textsubscript{2}~\cite{Safari-Appl.Surf.Sci.-464-153-2019,Tang-Sensors-21-1443-2021}.
 Wu \emph{et al},~\cite{Wu-RSCAdvances-14-1445-2024}
identified that by appending specific groups to arsenene, it becomes
extremely efficient in the location of molecules of CO, NO, NO\textsubscript{2}, SO\textsubscript{2}, NH\textsubscript{3}, and H\textsubscript{2}S. In addition, sensor materials capable of detecting gas molecules such as CO\textsubscript{2}, CH\textsubscript{4}, and N2~\cite{Oliveira-Sci.Rep.-12-22393-2022,Li-Mater.Chem.Phys.-301-127602-2023} have been investigated in the context of two-dimensional materials, including graphene.

Carbon nanotubes, specifically, demonstrate considerable promise in the realm of novel gas separation apparatus development~\cite{Poudel-Mater.TodayPhys.-7-7-2018,Cezar-arXiv-2023}. For adsorbing gases such as CO\textsubscript{2}~\cite{Alexiadis-Chem.Phys.Lett.-460-512-2008},
CH\textsubscript{4}~\cite{Lithoxoos-J.ofSupercriticalFluids-55-510-2010}, and H\textsubscript{2}~\cite{Dillon-Nature-386-377-1997,
Lyu-Nanomaterials-10-255-2020, Cezar-arXiv-2023}, single-walled carbon nanotubes (SWCNT) have been used. Furthermore, investigations into the electronic characteristics of carbon nanotubes after the adsorption of gas molecules (NO\textsubscript{2}, O\textsubscript{2}, NH\textsubscript{3}, N\textsubscript{2}, CO\textsubscript{2}, CH\textsubscript{4}, H\textsubscript{2}O, H\textsubscript{2}, and Ar) have shown that the electronic properties of SWCNTs are exceptionally susceptible to the adsorption of gases such as NO\textsubscript{2} and
O\textsubscript{2}~\cite{Cezar-arXiv-2023,Ganji-Commun.Theor.Phys.-53-987-2010}.  Lu \emph{et al}. In addition, the interactions between carbon nanobelt dimers (CNB) and small gas molecules (N\textsubscript{2}, CH\textsubscript{4}, CO, CO\textsubscript{2}, H\textsubscript{2}, H\textsubscript{2}O, H\textsubscript{2}O\textsubscript{2}, and O\textsubscript{2}). They discovered that these molecules tend to be adsorbed more strongly
inside the ring than outside, mainly because the interaction is much
stronger~\cite{Lu-Catalysts-12-561-2022}.

Ahmed \emph{et al}. conducted a recent study to assess the H\textsubscript{2}S gas detection capability utilizing Möbius carbon nanobelt strips (MCNBs), which demonstrated exceptional sensitivity to the gases under investigation. Furthermore, the presence of negative entropy values indicated that each of the formed complex structures exhibited thermodynamic stability. On the contrary, a more pronounced interaction with the CH\textsubscript{4} gas was observed in the
MCNB structure~\cite{Ahmed-R.Soc.OpenSci.-9-220778-2022}. Such theoretical investigations aid in establishing a correlation between the interaction properties and the structure of carbon nanomaterials.  In addition, they provide theoretical guidance for practical implementations that aid in the development of sensors for specific gases~\cite{Ahmed-R.Soc.OpenSci.-9-220778-2022}. Despite this, research on gas adsorption in various carbon structures is still
limited. Therefore, computational modeling can serve as a crucial instrument in understanding the underlying chemical and physical molecular mechanisms by furnishing pertinent data at the atomic level. When considering this, it is possible that properties obtained from diverse topologies could provide an alternative resolution to current issues such as air pollution resulting from chemical contaminants, which negatively impact both the environment and living organisms~\cite{Lyu-Nanomaterials-10-255-2020,
Ahmed-R.Soc.OpenSci.-9-220778-2022,Cezar-arXiv-2023}.

To advance the field of gas detection, it is imperative to establish
interdisciplinary collaboration. Collaboration among engineers, physicists, chemists, and material scientists is imperative for the synthesis of novel materials, the characterization of their properties, and their integration into operational devices. Furthermore, by actively involving technology developers and policymakers, the process of translating laboratory discoveries into
practical applications that protect the environment and public health can be facilitated.

Using semiempirical tight-binding theory, the interactions of phosgene, nitric oxide, methanol, methane, carbon monoxide, carbon dioxide, phosgene, hydrogen sulfide, ammonia, and nitrogen dioxide with boron-nitride nanobelts were investigated in this study. Various methods were employed to characterize the systems: identification of the optimal interaction region, geometry optimization, molecular dynamics, electronic property calculations, and topology studies.

\section{Materials and Methods}
\label{Sec:Method}
\begin{figure}[htpb]
\centering
\includegraphics[width=\textwidth]{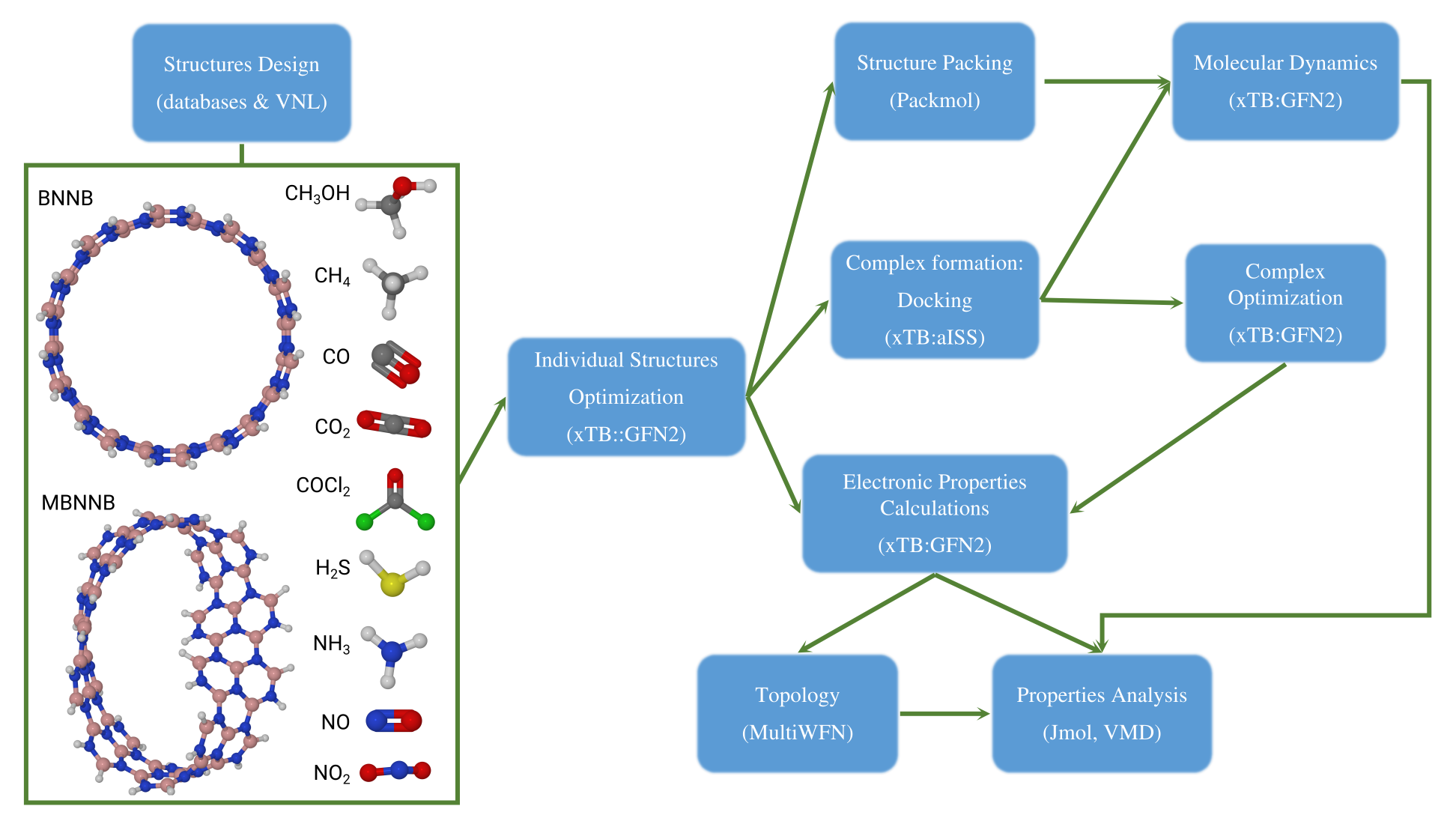}
\caption{Initial structures and flowchart of the used methodology.}
\label{Fig:StructMethod}
\end{figure}

This study employed two distinct nanostructures composed of boron nitride (BN): a nanobelt and a Möbius nanobelt, also known as a twisted nanobelt.  Beginning with two unit cells of a (10,0) boron-nitride nanosheet repeated ten times in the z direction, the Virtual NanoLab Atomistix Toolkit software~\cite{VNL} generated the nanobelt. Subsequently, the nanobelt was wrapped in a 360~degree motion, the periodicity was eliminated, and the border atoms were passivated
with hydrogen. In the case of Möbius nanobelt, after the initial repetition of cells, the nanobelt was twisted 180~degrees and then wrapped. The following nine greenhouse gases were used: nitrogen-dioxide (NO\textsubscript{2}), hydrogen-sulfide (H\textsubscript{2}S), methanol (CH\textsubscript{3}OH), methane (CH\textsubscript{4}), carbon monoxide (CO), carbon dioxide (CO\textsubscript{2}), phosgene (COCl\textsubscript{2}), ammonia (NH\textsubscript{3}), and nitrogen-oxide (NO). All structures were neutrally charged and are shown in
Figure~\ref{Fig:StructMethod}.

The systems are designated with the abbreviations BNNB and MBNNB to denote the boron-nitride nanobelt and MBNNB, respectively, and BNNB+gas (MBNNB+gas) to denote the complexes formed between BNNB (MBNNB) and greenhouse gases, such as BNNB+NO. The initial structures and flow diagram of the methodology used are shown in Figure~\ref{Fig:StructMethod}.

The calculations were conducted utilizing the implementation of the semiempirical tight binding method by the xTB program. The efficacy of the xTB package has been evaluated across multiple databases containing transition metals, organometallics, and lanthanide complexes. Compared to techniques such as coupled cluster and Density Functional Theory (DFT), the xTB package produced exceptionally accurate results~\cite{xTB_2,xTB_GFN0,xTB_GFN2,xTB_1}.

The calculations were carried out according to the sequence of procedures outlined in the methodology flow chart (refer to Figure~\ref{Fig:StructMethod}). Initially, the structures of each individual system (consisting of two nanobelts and nine gases) were optimized. Following this, the docking process is completed utilizing automated Interaction Site Screening (aISS)~\cite{xTB-dock}. Following a step that searches for pockets in molecule A (the nanobelts), a three-dimensional (3D) screening is performed to identify $\pi$-$\pi$ stacking interactions in various directions. Subsequently, approaches are made to identify the global orientations of molecule B (the gases) within an angular grid encircling molecule A.  In order to classify the
generated structures, the interaction energy (xTB-IFF) is utilized~\cite{xTB-IFF}. By default, for an additional two-step optimization of the genetic algorithm, one hundred structures with lower interaction energies are chosen. This ensures that conformations that were not detected in the initial screening are incorporated.
In the course of this two-step genetic optimization procedure, each pair of positions of molecule B is randomly crossed around molecule A. Subsequently, 50\% of the structures undergo random mutations in both position and angle. After ten iterations of this exhaustive search process, ten complexes with the lowest interaction energy are chosen. The structure with the lowest interaction energy is subsequently chosen as the input for the complex optimization.

The GFN2-xTB method, an exact self-consistent method that incorporates
multipole electrostatics and density-dependent dispersion contributions~\cite{xTB_GFN2}, was used to optimize the geometry in its
entirety. An extreme optimization level was ensured, with a convergence energy of $5\times10^{-8}$~E\textsubscript{h} and gradient norm convergence of $5\times10^{-5}$~E\textsubscript{h}/a\textsubscript{0} (where
a\textsubscript{0} is the Bohr radius).

The electronic properties were calculated using the spin polarization scheme and included the system energy, the highest occupied molecular orbital energy (HOMO, $\varepsilon_H$), the lowest unoccupied molecular orbital energy (LUMO, $\varepsilon_L$), the energy gap between the orbitals HOMO and LUMO ($\Delta \varepsilon=\varepsilon_H - \varepsilon_L$), and the electron transfer integrals.

The change in conductivity ($\Delta \sigma$) of a material leads to a
change in the electric signal and can be used to measure the sensitivity of a sensor~\cite{AbdalkareemJasim-Inorg.Chem.Commun.-146-110158-2022,Goel-EngineeringReports-5--2023}. These changes can be calculated as follows:

\begin{equation}
\label{Eq:Conductivity}
\Delta \sigma = (\sigma - \sigma_0)/\sigma_0,
\end{equation}
where $\sigma_0$ and $\sigma$ are the isolated nanobelts and complex
conductivities, respectively. As the conductivity of semiconductor material is proportional to the intrinsic carrier concentration (basically electrons) and their mobility, it is found that $\sigma$ is also proportional to the electronic gap as~\cite{Kittel1996}:

\begin{equation}
\label{Eq:ConductivityGap}
\sigma \propto Exp(-\Delta \varepsilon/2k_BT),
\end{equation}
$k_B$ and $T$ are the Boltzmann constant and system temperature, respectively. Higher values of $\Delta \sigma$ indicate a higher sensitivity of the nanobelt to the corresponding gas.

To study the electron / hole mobility between nanobelts and absorbed gas molecules, we determined the electron transfer integrals, also known as coupling integrals, using the dimer projection (DIPRO) method~\cite{xTB-DIPRO}. In this paper $J_{oc}$ represents the transport of the hole (occupied molecular orbitals), $J_{un}$
the transport of electrons (unoccupied molecular orbitals) and $J$ corresponds to the total charge transfer that includes the transport of hole / electrons between occupied / unoccupied molecular orbitals, respectively. Higher values of $J$ imply a higher coupling between the two fragments (i.e. a more intense interaction between them).

From the total system energies, the adsorption energy ($E_{ads}$) of the gas molecules adsorbed on the nanobelts was calculated using the following expression:

\begin{equation}
\label{Eq:bind}
E_{ads} = E_{NB+gas} - E_{NB}- E_{gas}.
\end{equation}

In equation~\ref{Eq:bind}, $E_{NB}$ and $E_{gas}$ are the energies
for the isolated nanobelts and the gas molecule, respectively, and
$E_{NB+gas}$ is the energy of the NB+gas complex (BNNB+gas and MNBB+gas
systems).

It should be noted that very strong interactions are not favorable for
gas detection because these imply that desorption of the adsorbate could be difficult and the device may suffer from long recovery times. The recovery time, one of the essential properties of gas-sensing materials, is exponentially related to the adsorption energy and was predicted using transition theory as follows:

\begin{equation}
\label{Eq:RecoveryTime}
\tau = \nu^{-1}_0 Exp (E_{ads}/k_BT)
\end{equation}
where $\nu_0$ is the exposed used frequency. In this study, we estimated the recovery time using $\nu_0$=10\textsuperscript{12}~Hz (corresponding to ultraviolet radiation) and T=298.15~K.

Using the wave function obtained when calculating the electronic properties, the topological properties and descriptors (such as critical points, electronic density, Laplacian of the electronic density, etc.) were determined using MULTIWFN~\cite{multiwfn} software.

As is well-known, geometry optimization involves using an algorithm to
obtain a local minimum structure on the potential energy surface (PES), which allows us to determine the lowest energy conformers of a system. However, this method does not provide information on the stability of the system over time. Molecular dynamics (MD) simulation, on the other hand, analyzes the movement of atoms and molecules at a specific temperature (here, 298.15~K) and provides a way to explore the PES.

\begin{figure}[htpb]
\centering
\includegraphics[width=8cm]{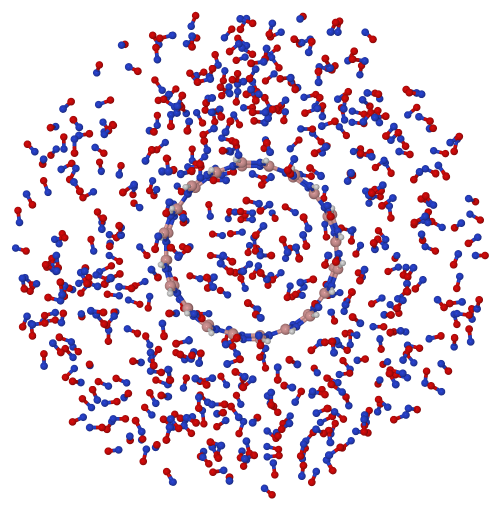}
\caption{A BNNB packaged with 500~NO molecules in a spherical conformation of 20~\AA~radii. Image rendered with Jmol software~\cite{jmol} using the CPK color scheme.}
\label{Fig:ComplexPackage}
\end{figure}

We perform molecular dynamics in two different cases. In the first case, we performed simulations on each complex, starting with the structures obtained from the aISS step as initial conformations. This gave an idea about how one single gas molecule interacts with each nanobelt. In the second case, using the PACKMOL software~\cite{packmol_0,packmol_1}, we created initial complexes,
adding 500~gas molecules to each nanobelt in a 20~\AA~radii spherical
distribution as shown in Figure~\ref{Fig:ComplexPackage}. This calculation gave information about how a more complex and realistic system evolved over time. Both molecular dynamics were run for a production time of 100~ps with a time step of 2~fs and an optional dump step of 50~fs, at which the final structure was written to a trajectory file. Molecular dynamics calculations were performed
using the GFN2-xTB method.

To characterize the particle distribution of a heterogeneous system, the radial distribution function (RDF) can be used. The RDF denoted in the equations by $g(r)$, is a pair correlation function that describes how, on average, the particles in a system are radially packed around each other~\cite{Hansen2006} and can be calculated using expression~\ref{Eq:RDF}:

\begin{equation}
\label{Eq:RDF}
g(\bf{r}) = \frac{n(\bf{r})}{\rho 4 \pi \bf{r}^2 \Delta \bf{r}},
\end{equation}
where $n(\bf{r})$ is the mean number of particles in a shell of width $\Delta \bf{r}$ at distance $\bf{r}$, and $\rho$ is the mean particle density.

The RDF ($g(\bf{r})$) not only describes the spatial correlation between two particles, but can also be used to describe the potential of the mean force~\cite{Hansen2006}. This property indicates how the free energy changes as a function of some inter- or intramolecular coordinate and can be calculated as

\begin{equation}
\label{Eq:PMF}
W(\bf{r}) = - k_B T ln[g(\bf{r})].
\end{equation}

\section{Results and discussion}
\label{Sec:results}
\subsection{Gas adsorption at the BN nanobelts}
\label{Sec:Geometry}
The calculated parameters for the optimized geometry complex are presented in Table~\ref{Tab:ResultsAdsBNNB} for the BNNB complexes and in Table~\ref{Tab:ResultsAdsMBNNB} for the MBNNB complexes. In both tables, the data are presented in increasing order of the adsorption energy (lower negative values imply better/stronger adsorption). At first sight, all of the gases presented negative adsorption energies, indicating a favorable interaction with both types of nanobelts. The NO, NH\textsubscript{3}, CH\textsubscript{3}OH, COCl\textsubscript{2}, and NO\textsubscript{2} gases are in the top-5 gases interacting in common for both systems, BNNB and MBNNB. However, a detailed analysis reveals that the Möbius boron-nitride-type nanobelt always presented lower values of the adsorption energies for all gases, ranging from 13\% (NH\textsubscript{3}) to 57\% (CH\textsubscript{4})
higher than the BNNB systems.

Even when the MBNNB has better adsorption energies for all gases than the BNNB, its sensitivity to each gas is different and is not aligned with the values of $E_{ads}$. Changes in material conductivity, $\Delta \sigma$, can be used to measure material sensitivity when interacting with another molecule. As described by Equation~\ref{Eq:Conductivity}, the conductivity of a material is directly related to the electronic gap ($\Delta \varepsilon$) of
the material. The reference conductivity, $\sigma_0$, is for isolated
nanobelts (without interacting with gases). In tables~\ref{Tab:ResultsAdsBNNB} and~\ref{Tab:ResultsAdsMBNNB}, both values for $\Delta \varepsilon$ and $\Delta \sigma$ are shown. Positive (negative) $\Delta \sigma$ represents an increase (decrease) in conductivity. The higher the value of $\Delta \sigma$, the more sensitive the sensor is. Methane, CH\textsubscript{4}, is the only gas that cannot be detected by either nanobelt, as its adsorption did not produce any change in the conductivity of the belt, for example, $\Delta \sigma=0$. However, dissimilar values of $\Delta \sigma$ are good
as they can be used as a parameter to measure sensor specificity, that is, given different electric responses for different adsorbed gases. Except for NH\textsubscript{3} and CH\textsubscript{3}OH, which showed very similar variations in conductivity for BNNB, both nanobelts can be used not only as adsorbents, but also to detect the presence of each of the gases studied here.

As pointed out before, if the sensor is intended to be reused, very strong interactions are not favorable because desorption of the adsorbate could be difficult. The reusability of the sensor can be related to the time it takes to recover, i.e., to remove the adsorbate. The sensor recovery time $\tau$ is directly dependent on the adsorption energy and can be calculated using Equation~\ref{Eq:RecoveryTime}). The lower values of $\tau$ indicate that the gas can easily be detached from the nanobelt. As the BNNB system has the highest $E_{ads}$, it presents the lowest values for $\tau$ ranging between a few seconds and a few nanoseconds. In the case of the MBNB, lower values of $E_{ads}$ implied the highest value of the recovery time, ranging from a thousand seconds to hundreds of nanoseconds. In all cases, the recoverable times are feasible.

\begin{table}[htpb]
\caption{Adsorption energy ($E_{ads}$), HOMO ($\varepsilon_H$), LUMO
($\varepsilon_L$), gap ($\Delta \varepsilon$), electrical conductivity variation ($\% \Delta \sigma$) by the adsorption, recovery time ($\tau$), and effective electron transfer integral ($J_{oc}$/$J_{un}$/$J$) for the BNNB systems$^\dagger$.}
\label{Tab:ResultsAdsBNNB}
\begin{center}
\setlength\extrarowheight{-3pt}
\begin{tabular}{lrrrrrrrr}
  \hline
  System & $E_{ads}$ & $\varepsilon_H$ &
  $\varepsilon_L$ & $\Delta \varepsilon$ & $\% \Delta \sigma$  & $\tau$ &
  $J_{oc}$/$J_{un}$/$J$ \\
  \hline
  \hline
  BNNB & --- & -9.503 & -5.530 & 3.973 & --- & --- & ---\\
  BNNB+NO & -0.768 & -8.113 & -5.759 & 2.354 & 292 & 16.46~s & ---\\
  BNNB+NH\textsubscript{3}   & -0.650 & -9.496 & -5.555 & 3.941 & 3 &
  0.15~s & 4/6/8\\
  BNNB+CH\textsubscript{3}OH & -0.481 & -9.503 & -5.559 & 3.944 & 2 &
  187.09~$\mu$s & 12/22/170 \\
  BNNB+COCl\textsubscript{2} & -0.476 & -9.519 & -8.071 & 1.447 & 743 &
  156.78~$\mu$s & 130/54/239\\
  BNNB+NO\textsubscript{2}   & -0.394 & -9.416 & -7.662 & 1.755 & 551 &
  5.95~$\mu$s & ---\\
  BNNB+H\textsubscript{2}S   & -0.284 & -9.496 & -5.595 & 3.901 & 6 &
  76.07~ns & 12/2/14\\
  BNNB+CO\textsubscript{2}   & -0.270 & -9.520 & -5.926 & 3.594 & 38 &
  44.38~ns & 6/23/21\\
  BNNB+CO& -0.261 & -9.519 & -6.852 & 2.667 & 201 &
  31.43~ns & 20/4/48\\
  BNNB+CH\textsubscript{4}   & -0.134 & -9.505 & -5.532 & 3.973 & 0 &
  0.20~ns & 11/8/17\\
  \hline
\end{tabular}
\begin{flushleft}
\tiny {$^\dagger$ $E_{b}$ is in units kcal/mol; $\varepsilon_H$,
$\varepsilon_L$, $\Delta \varepsilon$, are in units of $eV$ and $J_{oc}$,
$J_{un}$, and $J$ are in units of $meV$.}
\end{flushleft}
\end{center}
\end{table}

\begin{table}[htpb]
\caption{Adsorption energy ($E_{ads}$), HOMO ($\varepsilon_H$), LUMO
($\varepsilon_L$), gap ($\Delta \varepsilon$), electrical conductivity
variation ($\%\Delta \sigma$) by the adsorption, recovery time ($\tau$), and effective electron transfer integral ($J_{oc}$/$J_{un}$/$J$) for the MBNNB systems$^\dagger$.}
\label{Tab:ResultsAdsMBNNB}
\begin{center}
\setlength\extrarowheight{-3pt}
\begin{tabular}{lrrrrrrrr}
  \hline
  System & $E_{ads}$ & $\varepsilon_H$ &
  $\varepsilon_L$ & $\Delta \varepsilon$ & $\% \Delta \sigma$  & $\tau$ &
  $J_{oc}$/$J_{un}$/$J$ \\
  \hline
  \hline
  MBNNB & --- & -9.397 & -5.537 & 3.860 & --- & --- & ---\\
  MBNNB+NO& -0.919 & -8.271 & -5.595 & 2.676 &172 &
  6438.00~s & ---\\
  MBNNB+NH\textsubscript{3}   & -0.754 & -9.434 & -5.522 & 3.912 & -4 &
  9.20~s & 10/9/61\\
  MBNNB+COCl\textsubscript{2} & -0.654 & -9.411 & -8.068 & 1.344 & 737 &
  179.00~ms & 2/11/25\\
  MBNNB+CH\textsubscript{3}OH & -0.633 & -9.409 & -5.528 & 3.881 & -2 &
  76.23~ms & 9/49/19\\
  MBNNB+NO\textsubscript{2}   & -0.547 & -9.386 & -7.887 & 1.498 & 634 &
  2.55~ms & 3/68/81\\
  MBNNB+CO\textsubscript{2}   & -0.452 & -9.413 & -5.929 & 3.484 & 37 &
  58.54~$\mu$s & 6/74/66\\
  MBNNB+CO& -0.333 & -9.435 & -7.038 & 2.397 & 244 &
  537.45~ns & 8/90/120\\
  MBNNB+H\textsubscript{2}S   & -0.337 & -9.492 & -5.562 & 3.929 & -6 &
  635.86~ns & 3/22/20\\
  MBNNB+CH\textsubscript{4}   & -0.310 & -9.400 & -5.535 & 3.865 & 0 &
  214.32~ns & 30/19/49\\
  \hline
\end{tabular}
\begin{flushleft}
\tiny {$^\dagger$ $E_{b}$ is in units kcal/mol; $\varepsilon_H$,
$\varepsilon_L$, $\Delta \varepsilon$, are in units of $eV$ and $J_{oc}$,
$J_{un}$, and $J$ are in units of $meV$.}
\end{flushleft}
\end{center}
\end{table}

Herein, we refer to the set of the first three complexes with lower adsorption energy for each of the nanobelt types (BNNB and MBNNB) as the best-ranked complexes.

\subsection{Electronic properties}
\label{Sec:ElecProp}
Figures ~\ref{Fig:MO_BNNB} and~\ref{Fig:MO_MBNNB} show the calculated frontier orbitals HOMO (top row) and LUMO (bottom row) for the pristine nanobelts (without gases) and complexes of the highest ranked.

\renewcommand{\sizeA}{3.50cm}
\begin{figure}[tbph]
\centering
\begin{tabular}{cccc}
BNNB & BNNB+NO & BNNB+NH\textsubscript{3} & BNNB+CH\textsubscript{3}OH \\
\subfigure[]{\includegraphics[width=\sizeA]{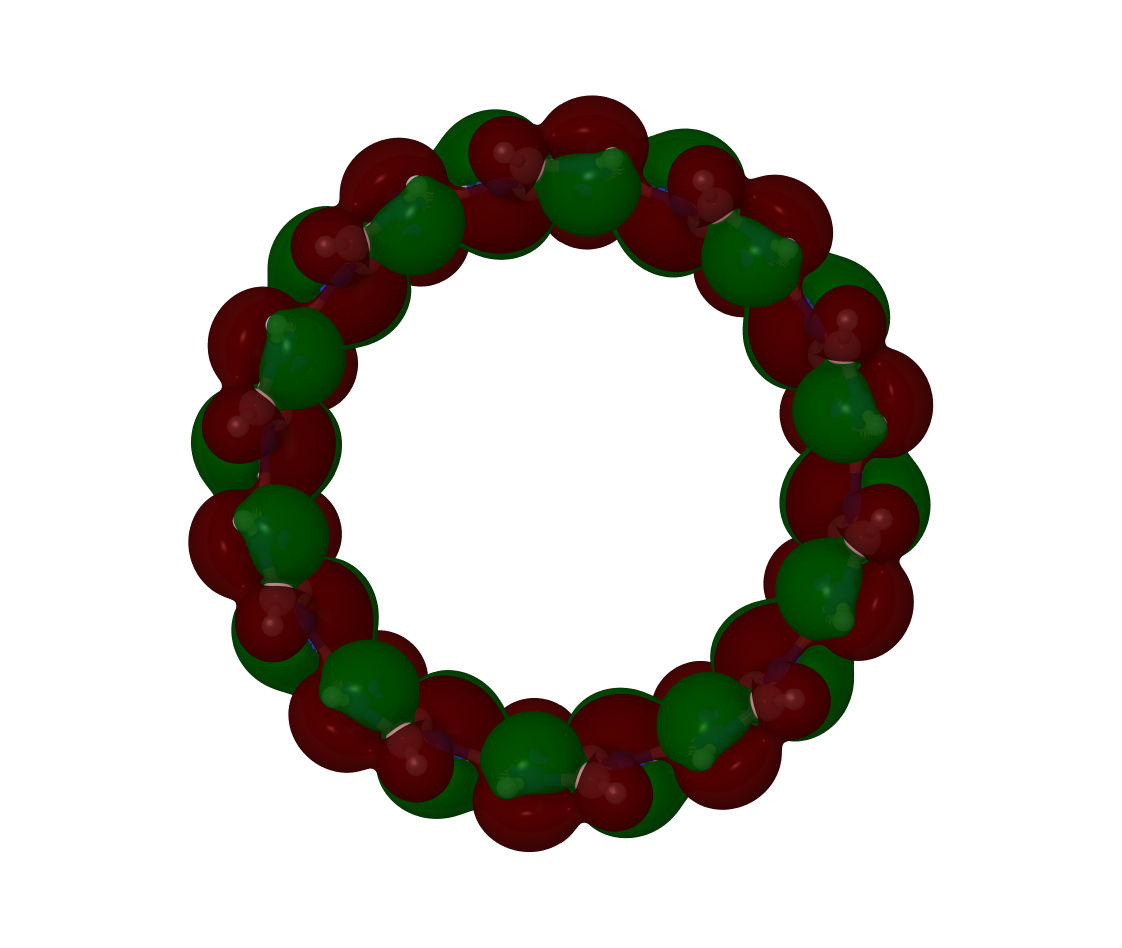}
\label{Fig:HOMO_BNNB}}  &
\subfigure[]{\includegraphics[width=\sizeA]{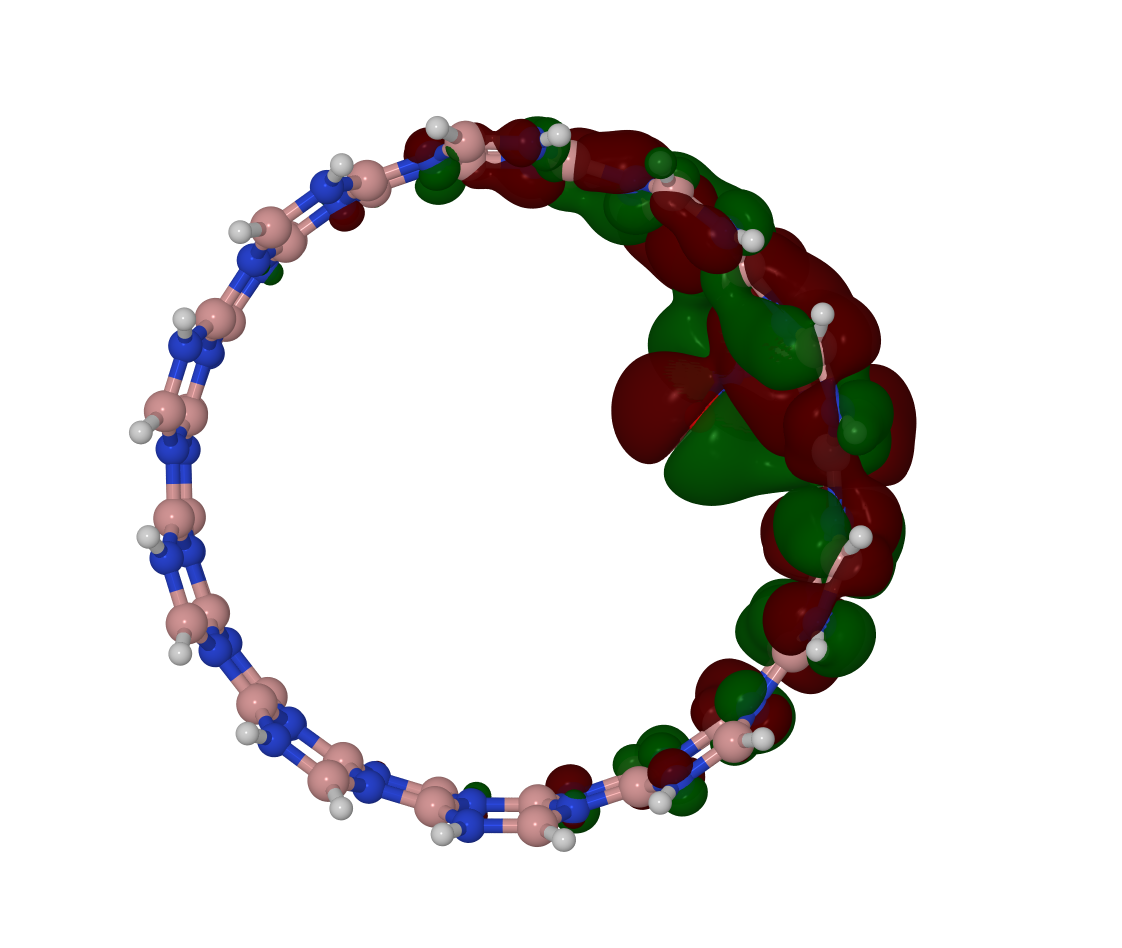}
\label{Fig:HOMO_BNNB+NO}}  &
\subfigure[]{\includegraphics[width=\sizeA]{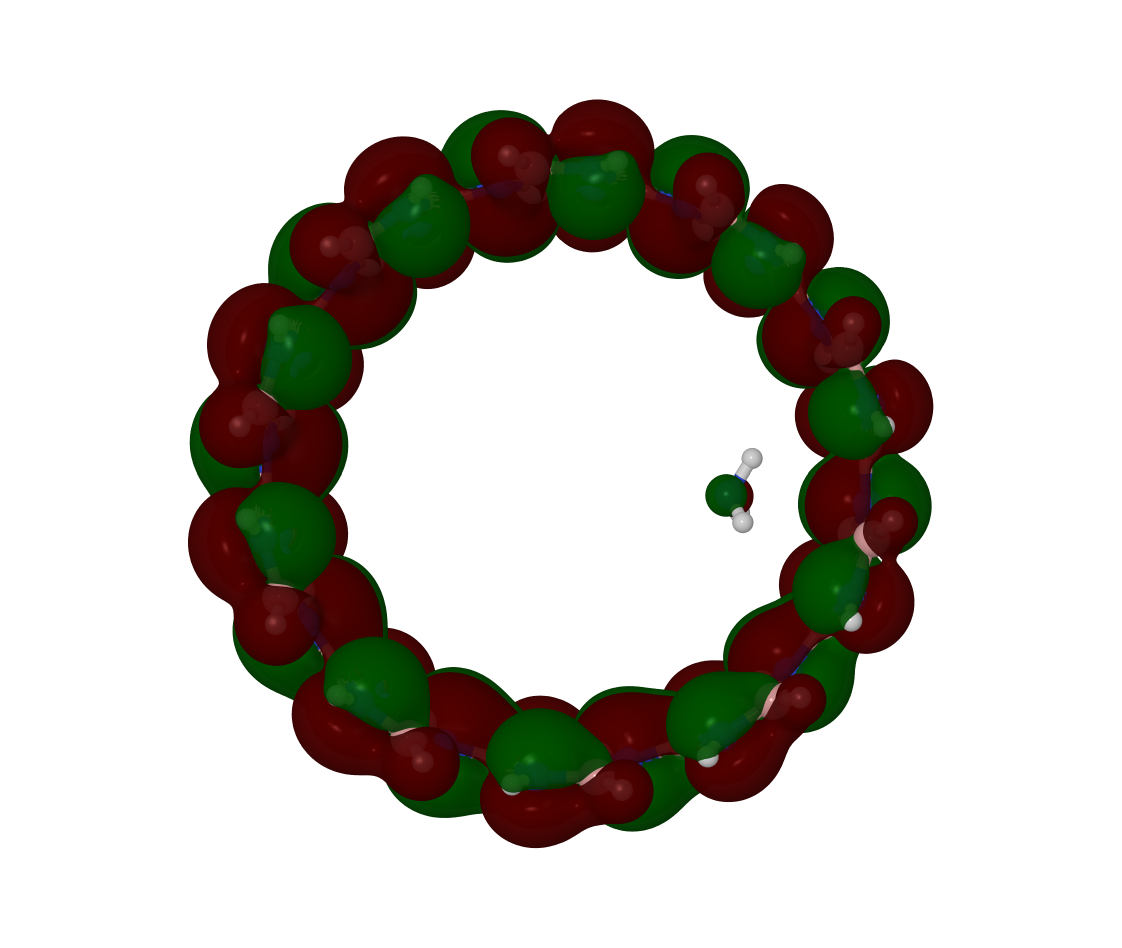}
\label{Fig:HOMO_BNNB+NH3}}  &
\subfigure[]{\includegraphics[width=\sizeA]{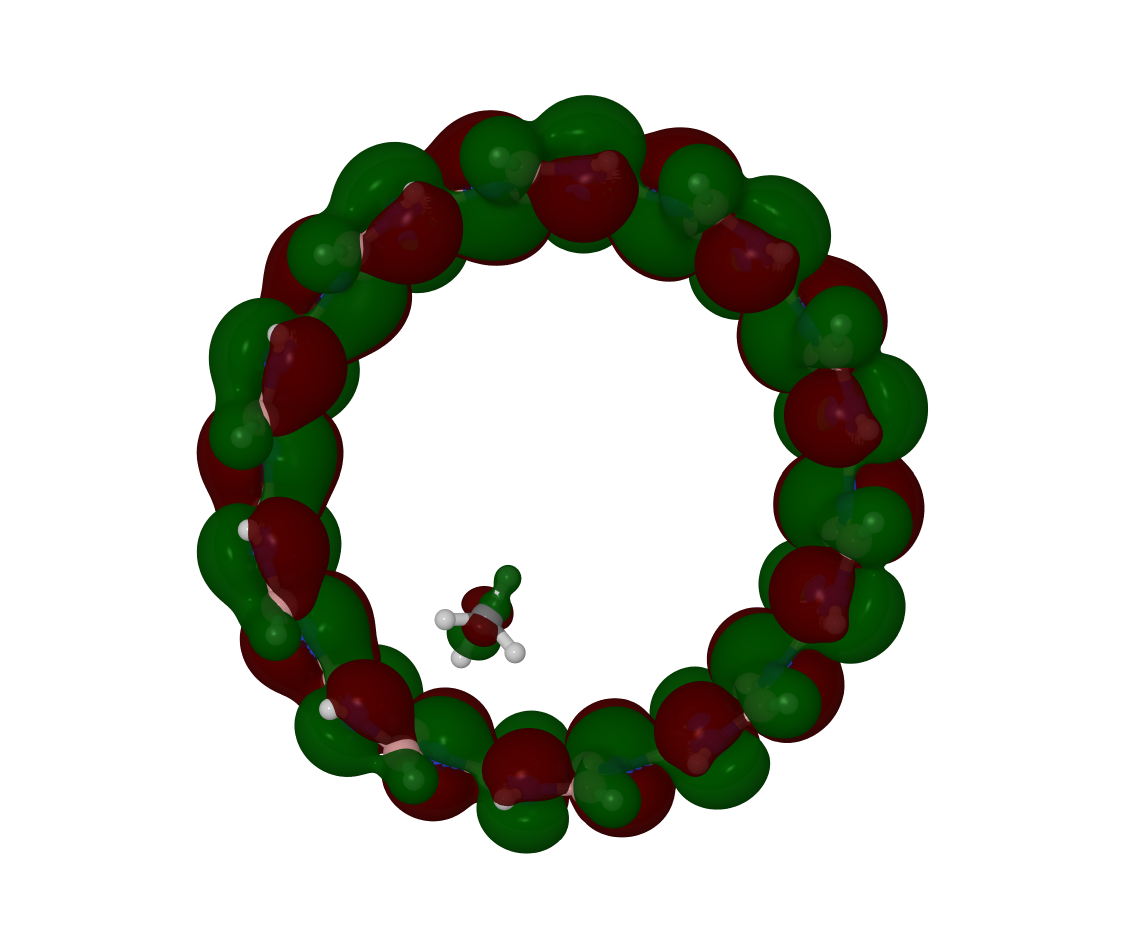}
\label{Fig:HOMO_BNNB+CH3OH}} \\

\subfigure[]{\includegraphics[width=\sizeA]{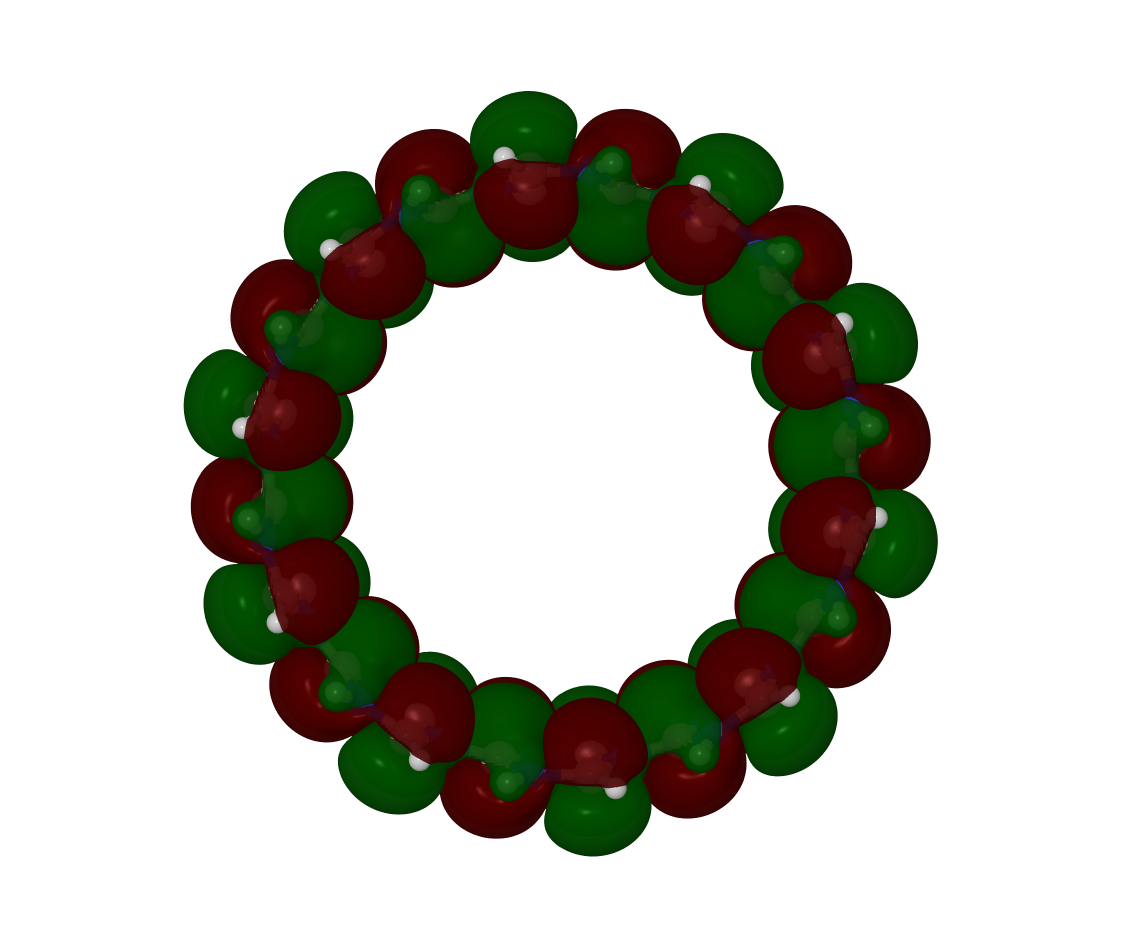}
\label{Fig:LUMO_BNNB}}  &
\subfigure[]{\includegraphics[width=\sizeA]{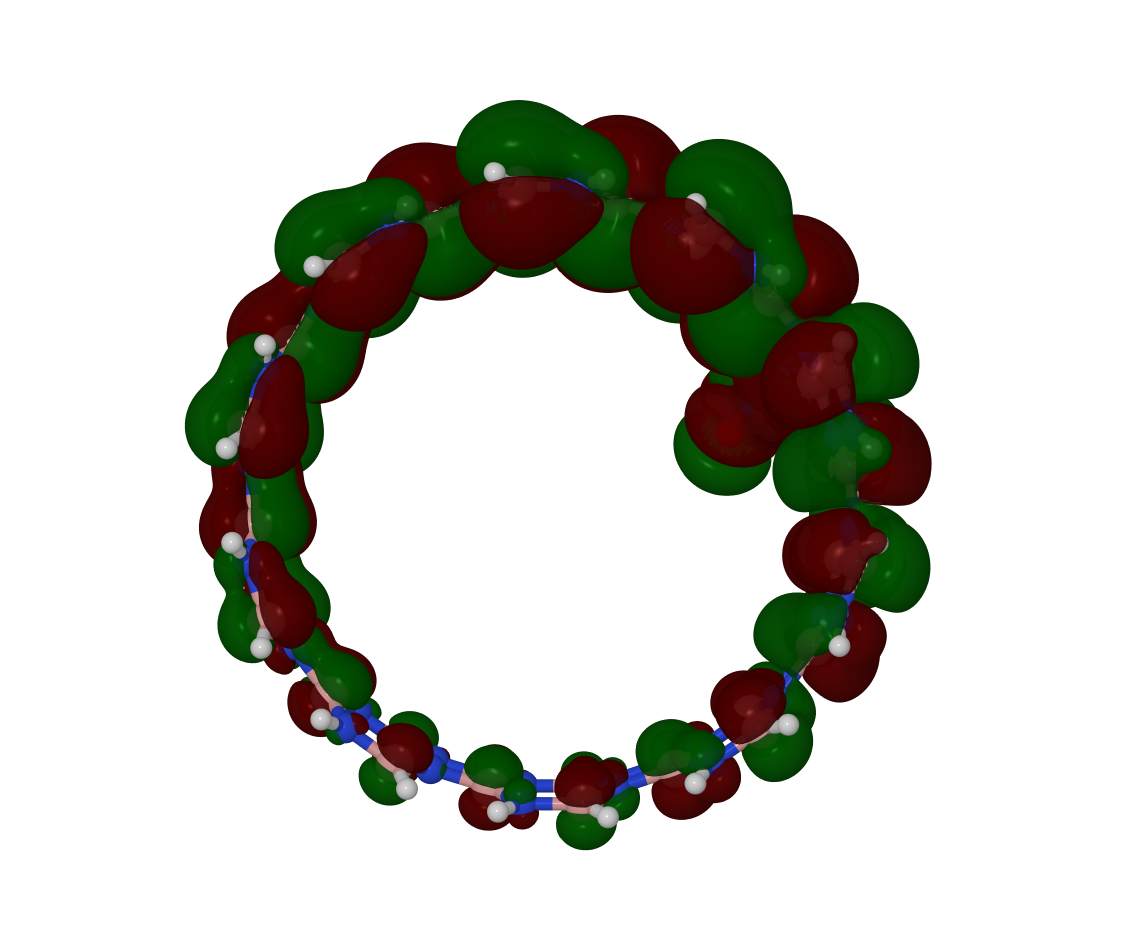}
\label{Fig:LUMO_BNNB+NO}}  &
\subfigure[]{\includegraphics[width=\sizeA]{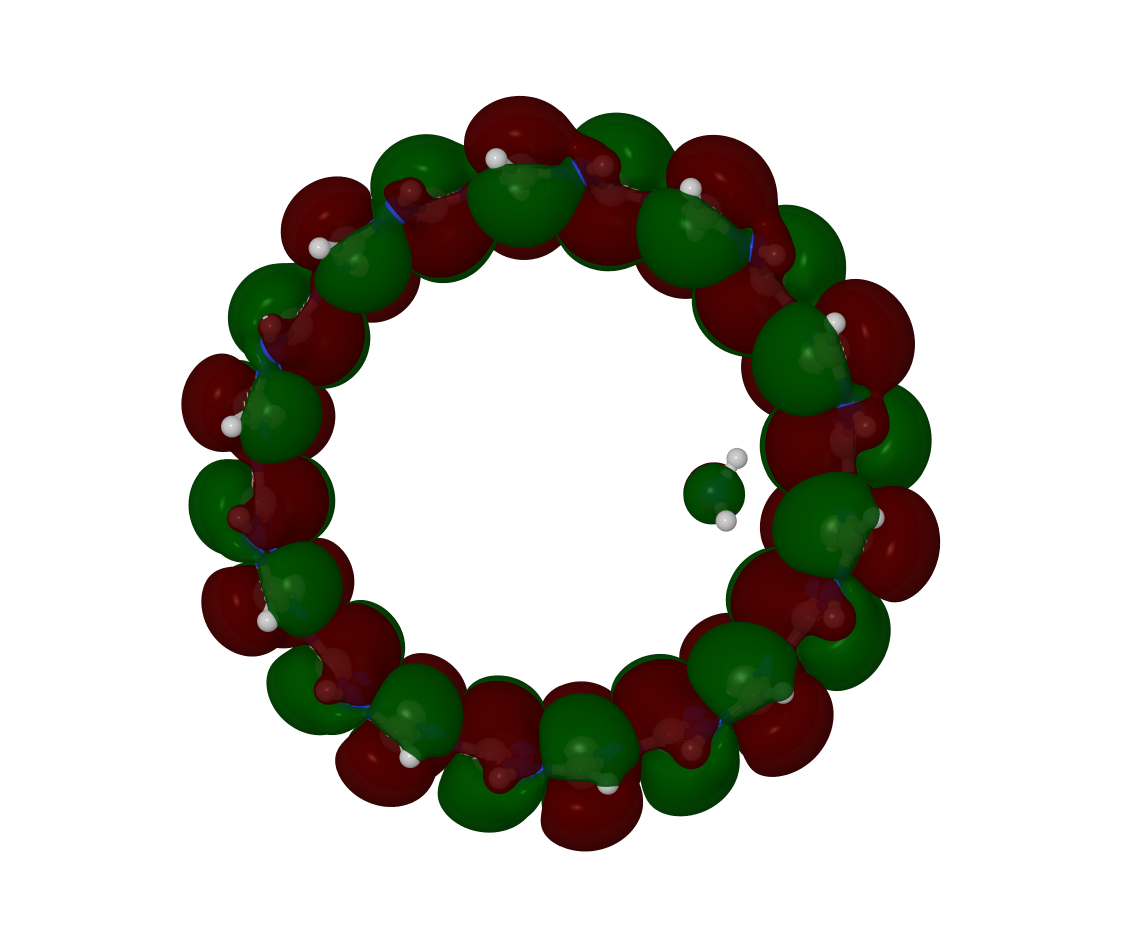}
\label{Fig:LUMO_BNNB+NH3}}  &
\subfigure[]{\includegraphics[width=\sizeA]{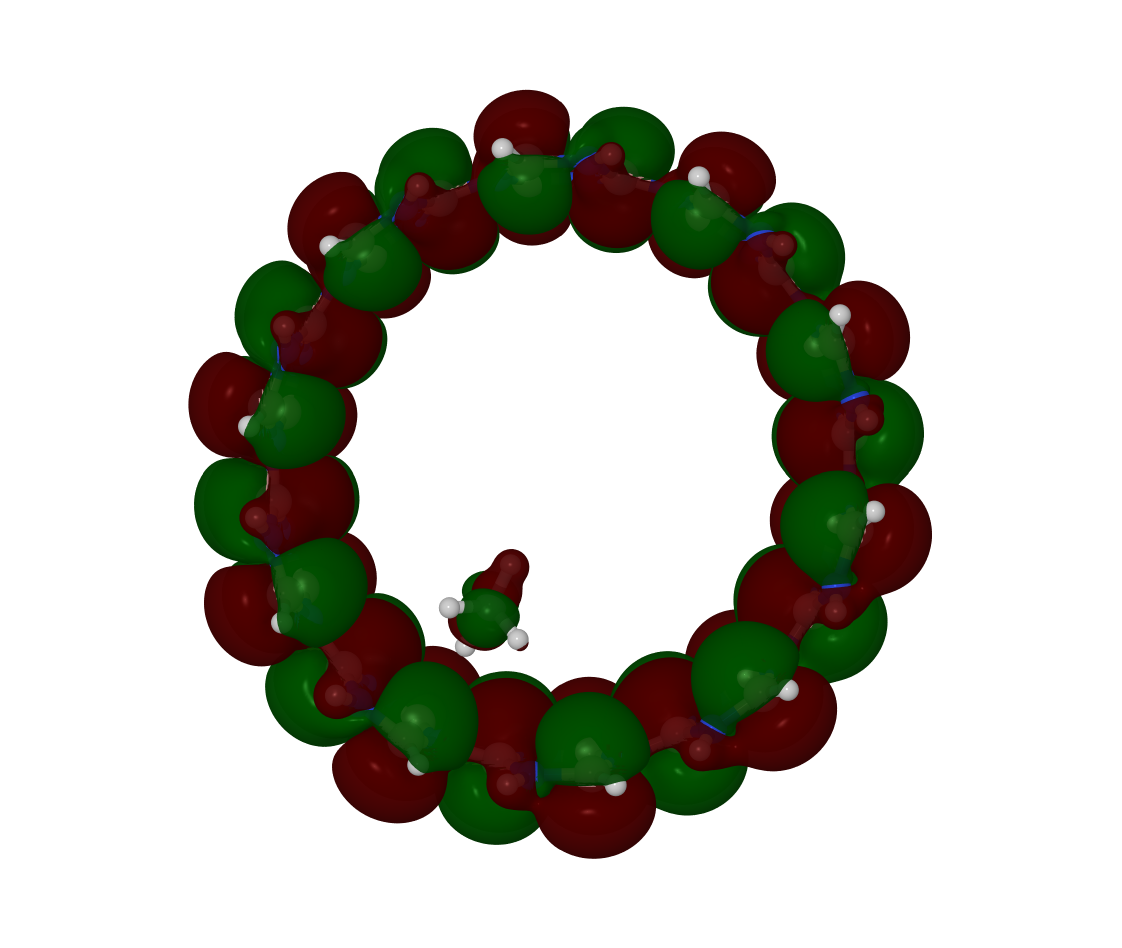}
\label{Fig:LUMO_BNNB+CH3OH}} \\
\end{tabular}
\caption{\label{Fig:MO_BNNB} HOMO (top row) and LUMO (bottom row) for the BNNB complexes and the most ranked. The red (green) represents negative (positive) values. Orbital surfaces were rendered with an isovalue equal to 0.001 and with Jmol software~\cite{jmol} using the CPK color scheme for atoms.}
\end{figure}

The BNNB, because of its symmetry, presents a homogeneous molecular orbital's distribution. Depending on the strength of the interaction with the gases, this distribution can be modified. The strongest interaction between NO and BNNB is reflected in the redistribution of the belt wave functions, now concentrated around the region of interaction with the NO gas. In this case, the DIPRO methodology detected only one fragment, indicating the formation of a covalent bond between BNNB and NO. This is confirmed by the topological results presented in Section~\ref{Sec:Topo}.

The other two gases, NH\textsubscript{3} and CH\textsubscript{3}OH, even when bound to the BNNB, induce slight modifications in the orbital surface distribution. In the case of BNNB+NH\textsubscript{3}, the calculated values of the effective electron transfer integral are very similar, which is in accordance with the surface redistribution of the small surfaces for both orbitals, HOMO and LUMO. The system BNNB+CH\textsubscript{3}OH also shows small surface modifications, but, in accordance with the values of the effective electron transfer integral, the LUMO surfaces experiment with slightly higher charge redistribution, which is reflected on the BNNB and H\textsubscript{3}OH LUMO surfaces.

\renewcommand{\sizeA}{3.50cm}
\begin{figure}[tbph]
\centering
\begin{tabular}{cccc}
MBNNB & MBNNB+NO & MBNNB+NH\textsubscript{3} & MBNNB+COCl\textsubscript{2} \\
\subfigure[]{\includegraphics[width=\sizeA]{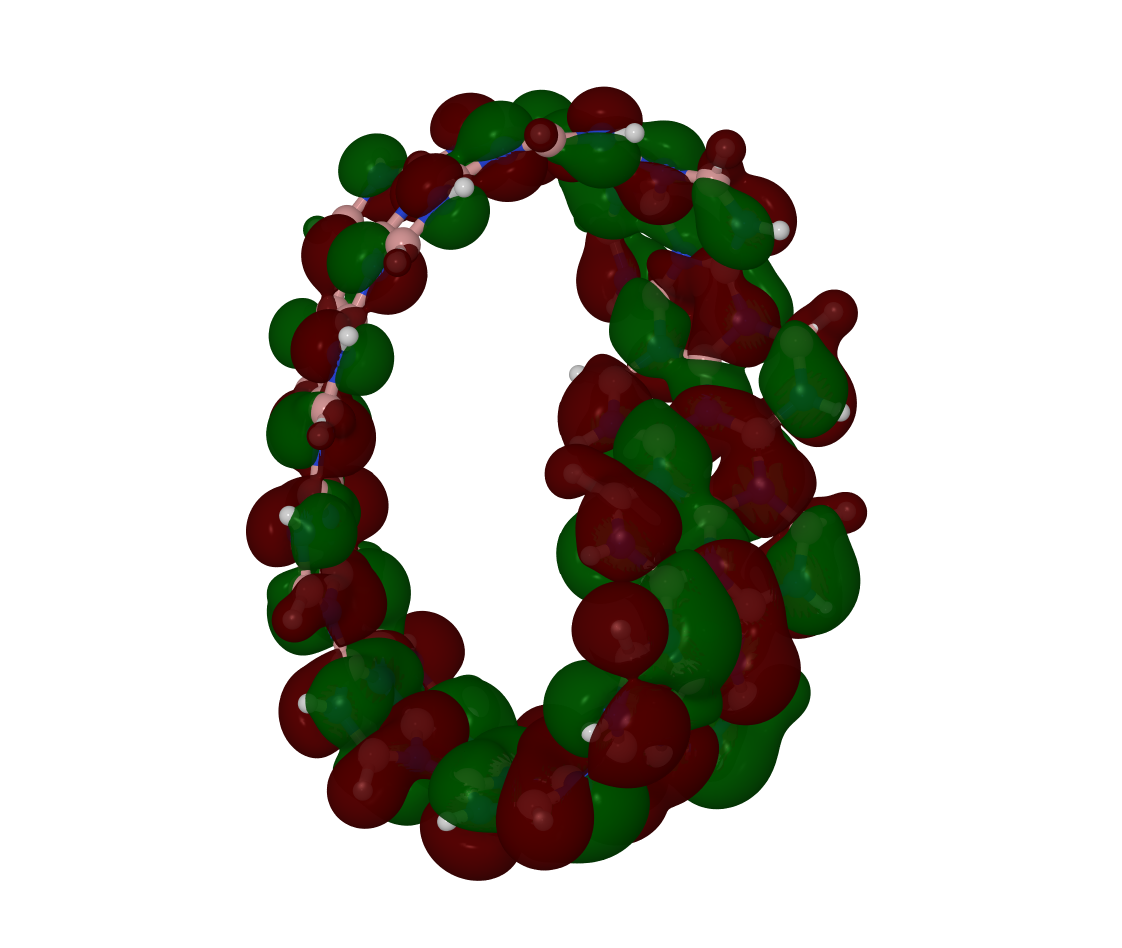}
\label{Fig:HOMO_MBNNB}}  &
\subfigure[]{\includegraphics[width=\sizeA]{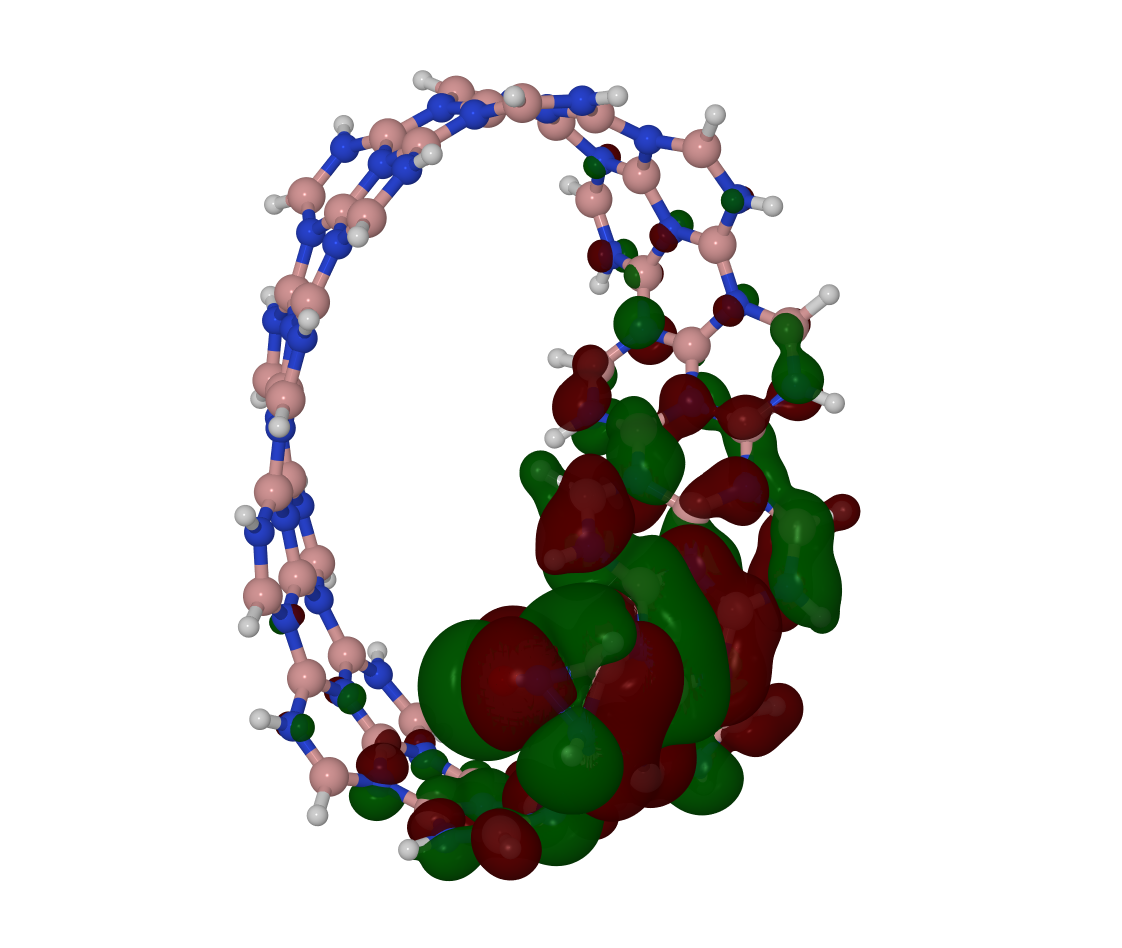}
\label{Fig:HOMO_MBNNB+NO}}  &
\subfigure[]{\includegraphics[width=\sizeA]{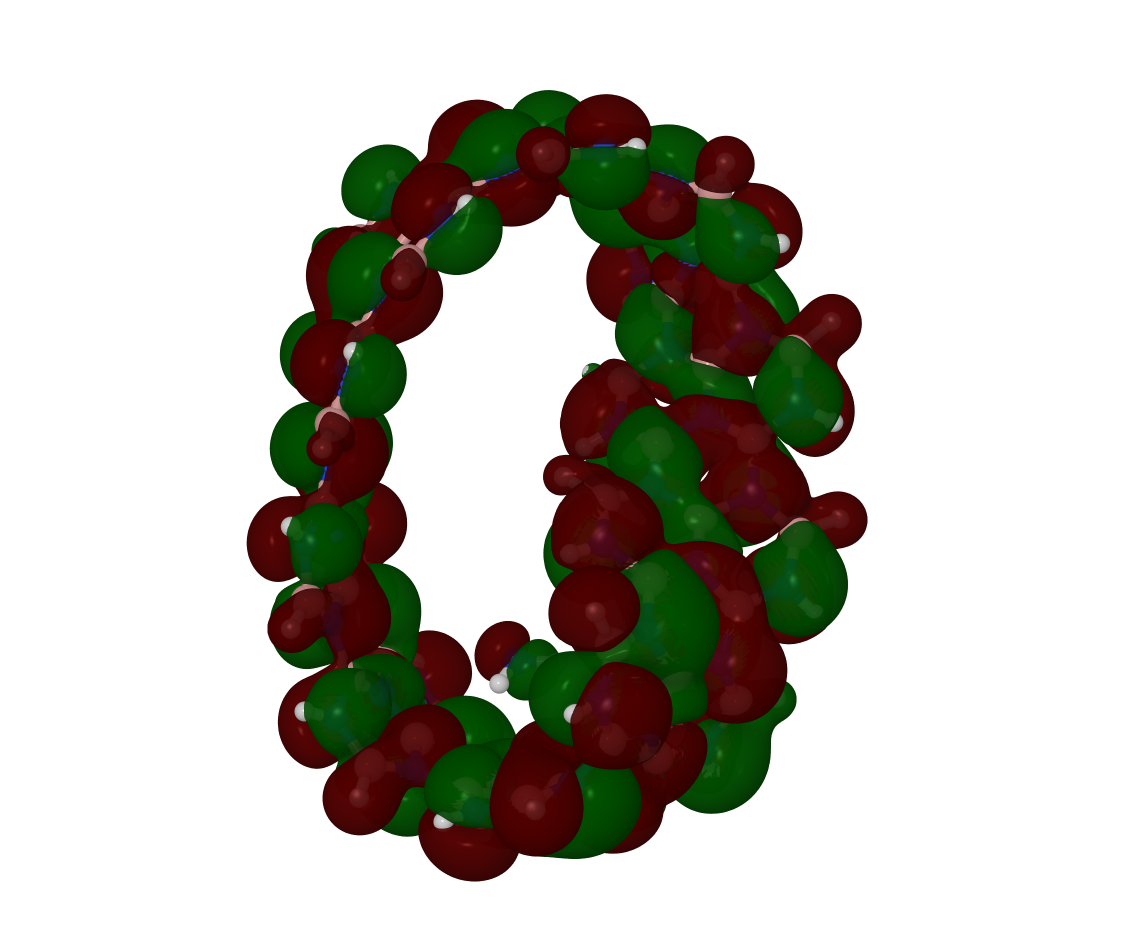}
\label{Fig:HOMO_MBNNB+NH3}}  &
\subfigure[]{\includegraphics[width=2.0cm]{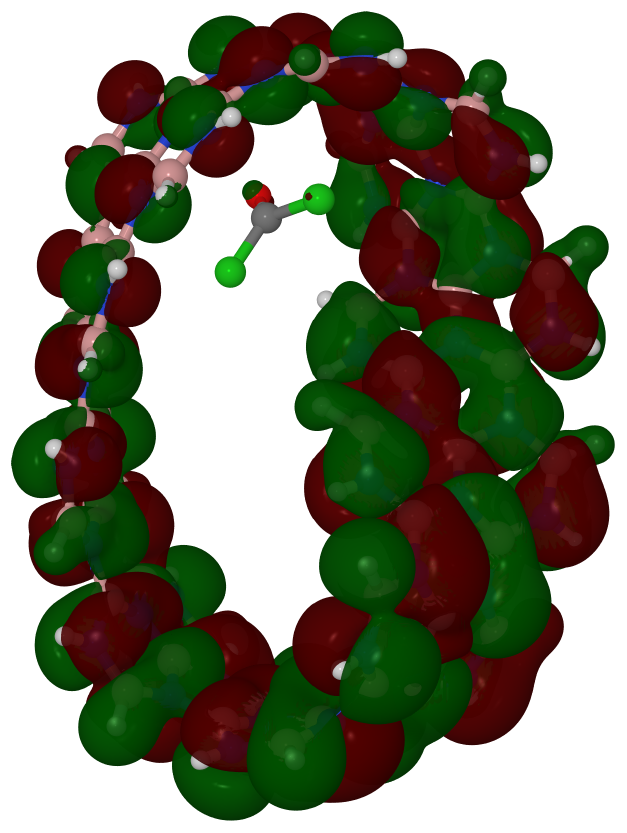}
\label{Fig:HOMO_MBNNB+COCl2}} \\

\subfigure[]{\includegraphics[width=\sizeA]{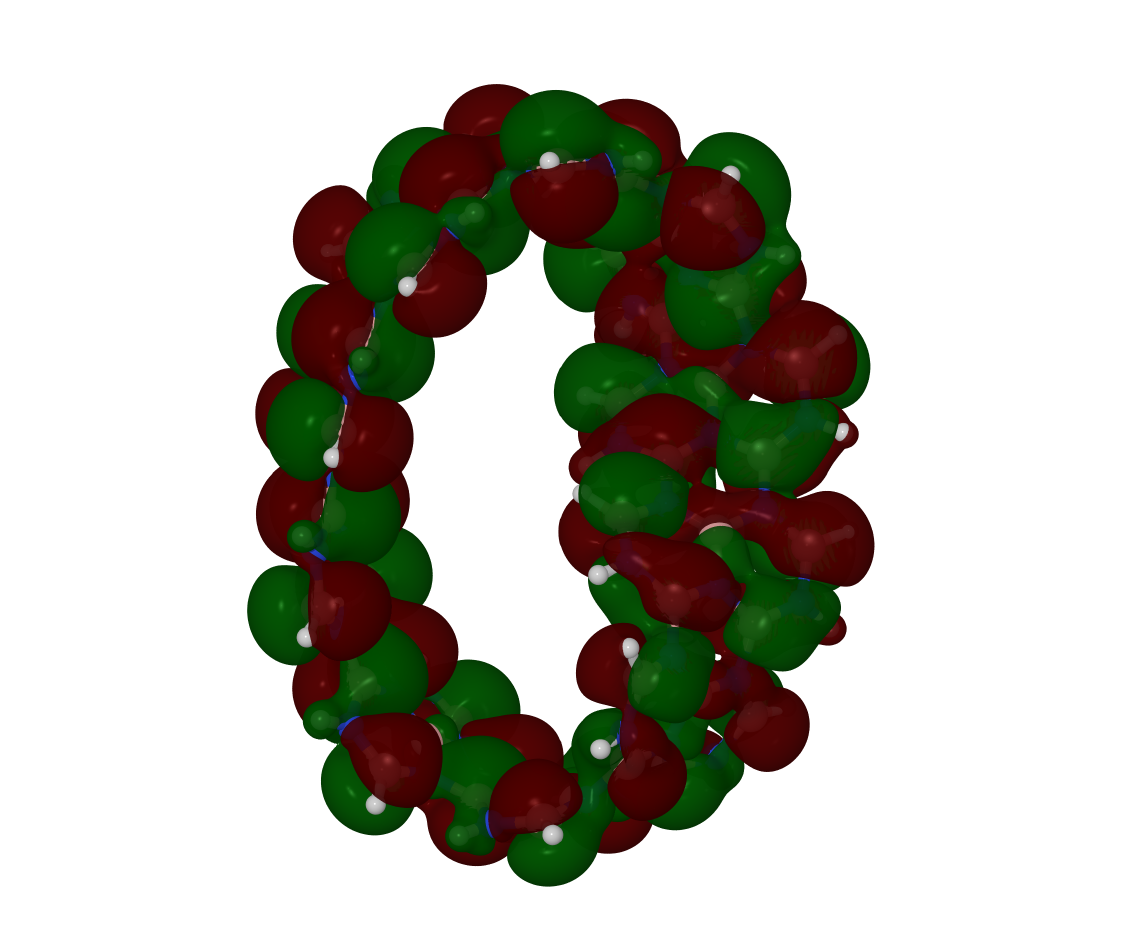}
\label{Fig:LUMO_MBNNB}}  &
\subfigure[]{\includegraphics[width=\sizeA]{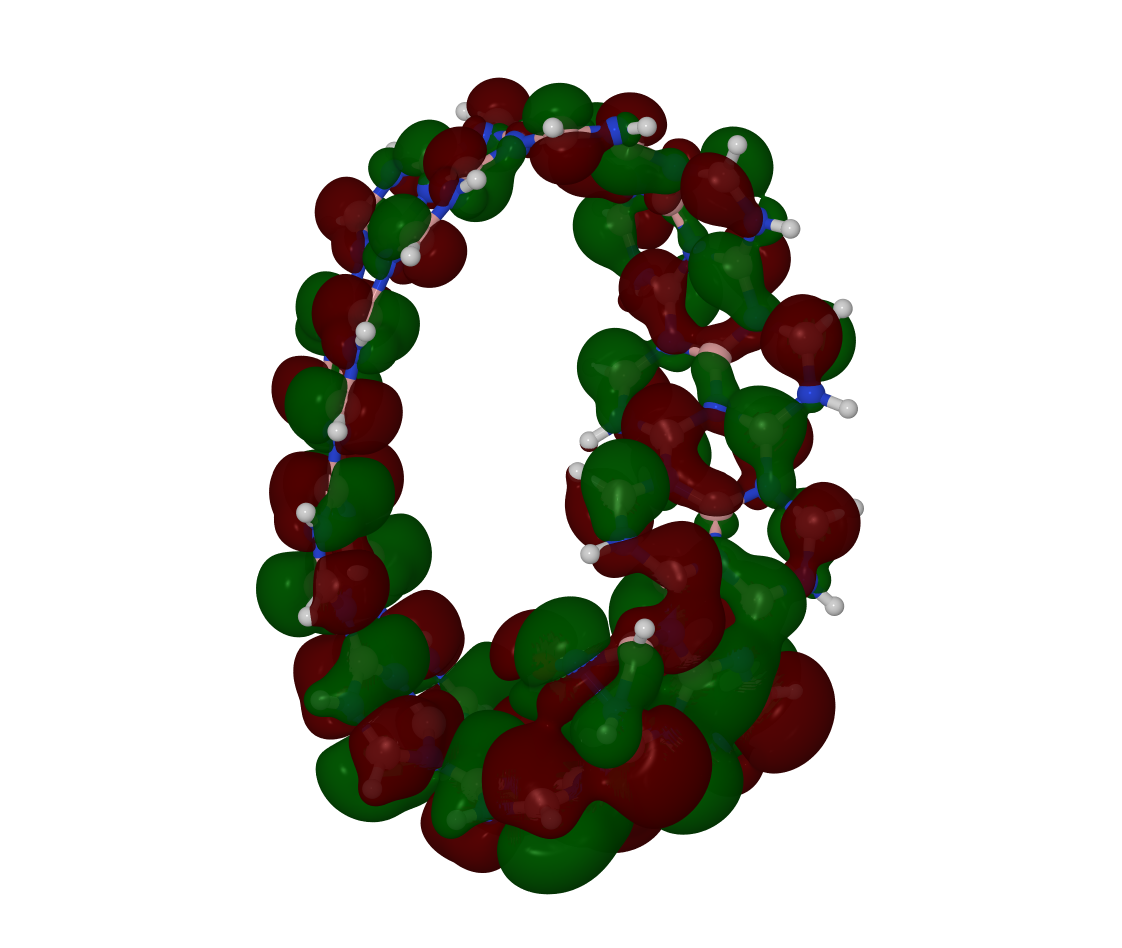}
\label{Fig:LUMO_MBNNB+NO}}  &
\subfigure[]{\includegraphics[width=\sizeA]{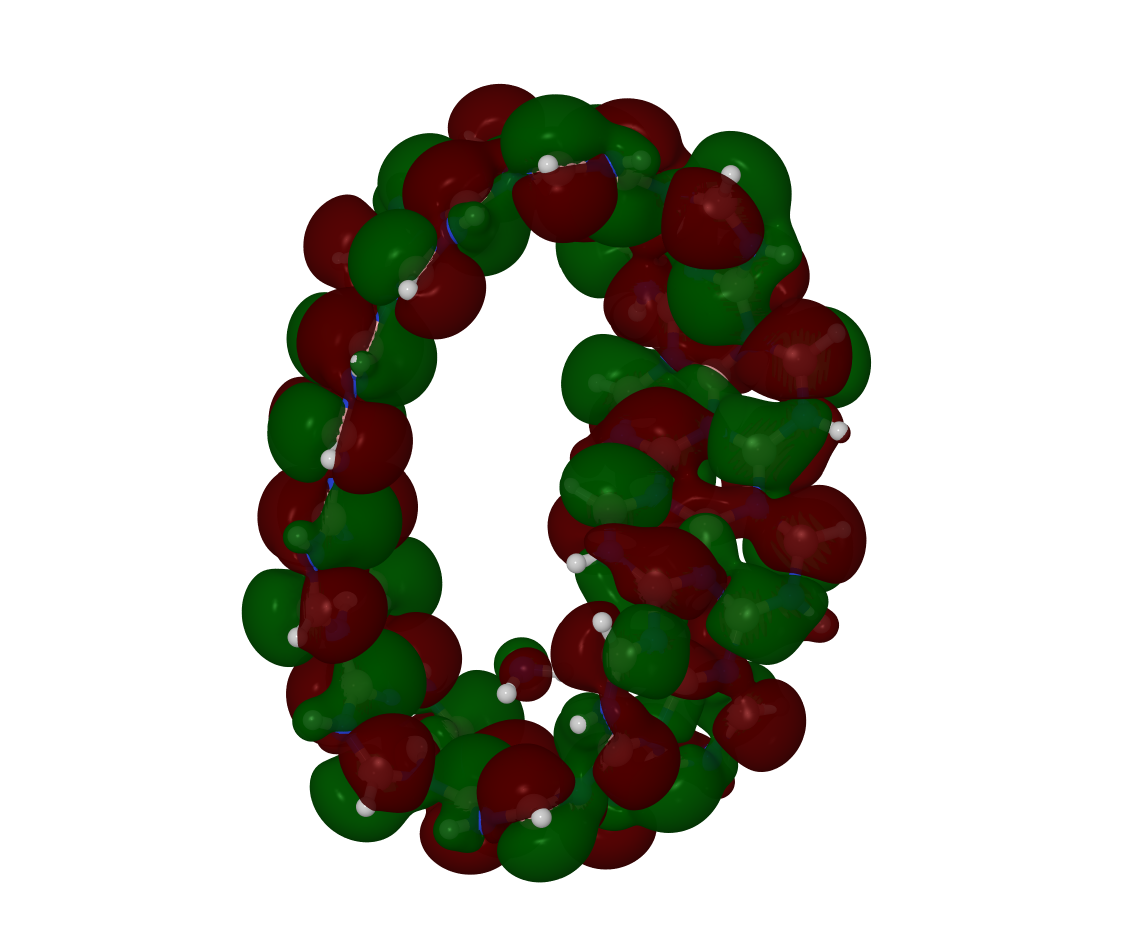}
\label{Fig:LUMO_MBNNB+NH3}}  &
\subfigure[]{\includegraphics[width=2.0cm]{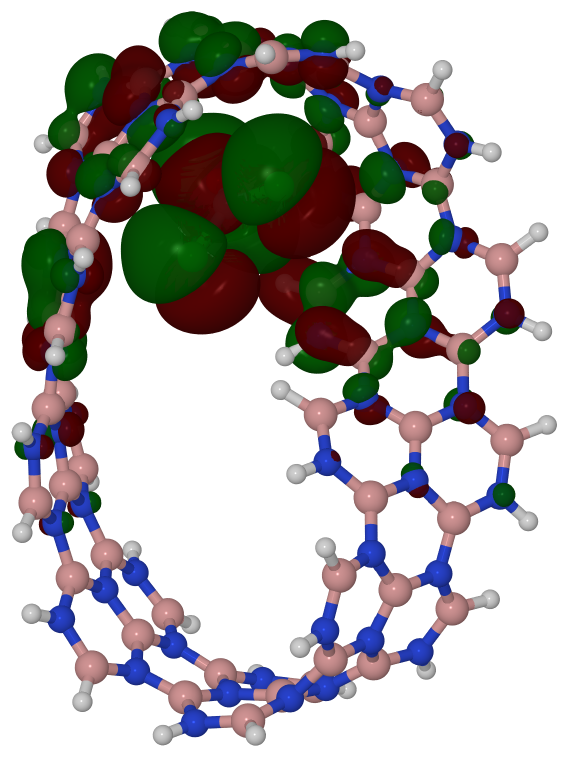}
\label{Fig:LUMO_MBNNB+COCl2}} \\
\end{tabular}
\caption{\label{Fig:MO_MBNNB} HOMO (top row) and LUMO (bottom row) for
MBNNB and the best-ranked complexes. Red (green) represents negative (positive) values. Orbital surfaces were rendered with an isovalue equal to 0.001 and with Jmol software~\cite{jmol} using the CPK color scheme for atoms.}
\end{figure}

The Möbius nanobelt, because of the induced twist, broke the frontier orbital distribution. In this case, there is a higher density around the twisted region, and the HOMO is more affected than the LUMO. The high interaction of MBNNB with NO is reflected in the HOMO/LUMO redistribution, localized around the NO adsorption region similar to the BNNB system. In this case, the DIPRO methodology detected that the system had only one fragment, i.e., covalently bonded.

 NH\textsubscript{3} bound to the MBNNB produces a small redistribution of the frontier orbital surfaces, which is in accordance with similar values for the effective electron transfer integral. However, even when the interaction of COCl\textsubscript{2} with MBNNB induced small modifications in the HOMO distribution, the surface of LUMO was highly modified, with the surface redistributed around the COCl\textsubscript{2} adsorption region. The latter is reflected in the value of $J_{un}$ (related to LUMO) that is more
than 5 times higher than the value of $J_{oc}$ (related to HOMO). The types of interactions between MBNNB and COCl\textsubscript{2} are discussed in Section~\ref{Sec:Topo}.

\subsection{Topological analysis}
\label{Sec:Topo}

The aim of topological analysis is to detect critical points, which are
positions where the gradient norm of the electron density value equals zero. Critical points are classified into four types based on negative
eigenvalues of the Hessian matrix of the real function~\cite{Bader1994}. The criteria for bond classification and types of critical points are described in detail in ref.~\cite{myMethods}.

The use of the electron density ($\rho$), the Laplacian of the electron density ($\nabla ^2\rho$), the electron localization function (ELF) index, and the localized orbital locator (LOL) index provides insight into the bond type (covalent or noncovalent) in various systems. The location of the electron movement is related to the ELF index, which ranges from 0 to 1~\cite{elf,elf2}. High values of the ELF index indicate a high degree of electron localization, suggesting the presence of a covalent bond. LOL index is another function that can be used to identify regions of high localization~\cite{lol}. The LOL
index also ranges from 0 to 1, with smaller (larger) values usually occurring in the boundary (inner) regions.

\renewcommand{\sizeA}{3.50cm}
\begin{figure}[tbph]
\centering
\begin{tabular}{ccc}
\subfigure[BNNB+NO]{\includegraphics[width=\sizeA]{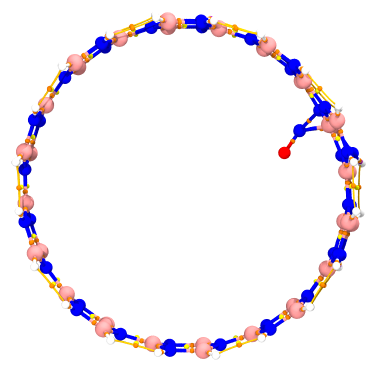}
\label{Fig:TOPO_BNNB+NO}}  &
\subfigure[BNNB+NH\textsubscript{3}]{\includegraphics[width=\sizeA]{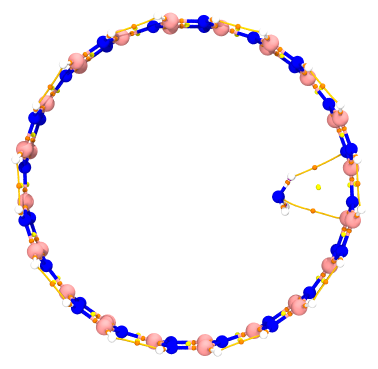}
\label{Fig:TOPO_BNNB+NH3}}  &
\subfigure[BNNB+CH\textsubscript{3}OH]{\includegraphics[width=3.35cm]{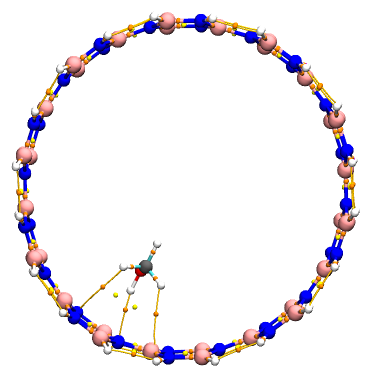}
\label{Fig:TOPO_BNNB+CH3OH}} \\

\subfigure[MBNNB+NO]{\includegraphics[width=\sizeA]{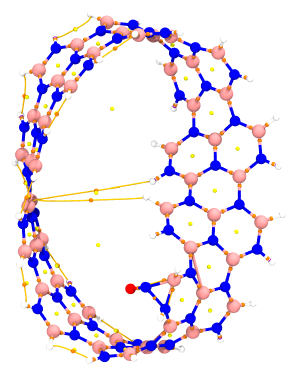}
\label{Fig:TOPO_MBNNB+NO}}  &
\subfigure[MBNNB+NH\textsubscript{3}]{\includegraphics[width=\sizeA]{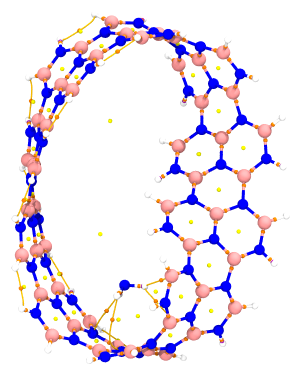}
\label{Fig:TOPO_MBNNB+NH3}}  &
\subfigure[MBNNB+COCl\textsubscript{2}]{\includegraphics[width=3.35cm]{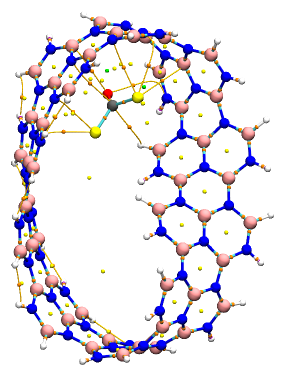}
\label{Fig:TOPO_MBNNB+COCl2}} \\
\end{tabular}
\caption{\label{Fig:3D-CPs} Critical points: BCPs, RCPs, and CCPs are
represented as orange, yellow, and green dots, respectively. Image rendered with VMD software~\cite{vmd} using the CPK color scheme for atoms.}
\end{figure}

The calculated critical points for the best ranked complexes are shown in Figure~\ref{Fig:3D-CPs}. Bond critical points (BCPs) are represented by orange dots, ring critical points (RCPs) by yellow dots, and cage critical points (CCPs) are represented by green dots. In the case of BCPs, the calculated descriptor properties are presented in Table~\ref{Tab:ResultsBCPs}. In addition to the BCPs, RCPs and CCPs are also formed. This implies that the electronic cloud of the belts is modified in a way favorable to the formation of the complexes.

For NO complexes, all descriptors indicate the formation of two covalent bonds with both types of nanobelts, respectively. In both cases, the nitrogen atom from NO is responsible for interacting with boron, and the nitrogen atoms from nanobelts. Higher values for MBNNB+NO’s descriptors imply that the interactions between NO and MBNNB are stronger than with BNNB. This was expected because the adsorption energies are higher for MBNNB+NO than for BNNB+NO (see Tables~\ref{Tab:ResultsAdsBNNB} and~\ref{Tab:ResultsAdsMBNNB}). An implication of the interaction between NO and MBNNB is the closure of the Möbius nanobelt and the formation of two new bonds from the atoms in the twisted region with the opposite ones (Figure~\ref{Fig:TOPO_MBNNB+NO}). This means that the induced twist created a type of two-pocket pocket that gave the Möbius nanobelt more
structural flexibility than the nontwisted belt has. This phenomenon was already observed for Möbius nanobelts based on boron nitride and carbon that interact with metal nanoclusters~\cite{Aguiar-Phys.BCondens.Matter-668-415178-2023, Aguiar-J.Mol.Model.-29-277-2023}.

The lower values of the descriptors for other complexes (Table~\ref{Tab:ResultsBCPs}), suggest that the bonds formed are noncovalent. A close look at the figures~\ref{Fig:TOPO_BNNB+NH3}, \ref{Fig:TOPO_BNNB+CH3OH}, \ref{Fig:TOPO_MBNNB+NH3}, and \ref{Fig:TOPO_MBNNB+COCl2}, shows that for the other gases, the interactions are mainly with hydrogen and chlorine gas atoms. It
is well known that hydrogen bond formation is weak and easily breakable
compared to covalent bonds~\cite{Jeffrey1997,Desiraju1999}. The latter is reflected in the lower recovery times ($\tau$) shown in Tables~\ref{Tab:ResultsAdsBNNB} and~\ref{Tab:ResultsAdsMBNNB}. A case of interest is the MBNNB + COCl complex\textsubscript{2} where 9 BCPs were found. In one bond, the hydrogen from the nanobelt is involved (BCP no. 301) and in the other five, the chlorine atom of the gas is involved. In this case, even when chlorine has an electronegativity similar to that of nitrogen (3.16 for Cl and 3.04 for N), the atomic radius of chlorine is higher, causing lower polarization and, in turn, weaker interactions than that of nitrogen.

\begin{table}[htpb]
\caption{Bond critical points, BCPs, and the atoms involved (nanobelt-gas),
$\rho$, $\nabla ^2\rho$, ELF and LOL
descriptors values$^\dagger$.}
\label{Tab:ResultsBCPs}
\begin{center}
\setlength\extrarowheight{-6pt}
\begin{tabular}{lrrrrr}
  \hline
  System  & BCPs (atoms) & $\rho$ & $\nabla ^2\rho$ & ELF &
  LOL\\
  \hline
  \hline
  BNNB+NO
& 310 (N-N) & 0.2099 & 0.0853 & 0.7300 & 0.6219 \\
& 309 (B-N) & 0.1322 & 0.3923 & 0.5650 & 0.5137 \\

  BNNB+NH\textsubscript{3}
& 286 (N-H) & 0.0054 & 0.0312 & 0.0074 & 0.0795 \\
& 259 (B-N) & 0.0051 & 0.0260 & 0.0084 & 0.0845 \\

  BNNB+CH\textsubscript{3}OH
&  35 (N-O) & 0.0125 & 0.0607 & 0.0242 & 0.1362 \\
& 196 (B-H) & 0.0033 & 0.0160 & 0.0055 & 0.0697 \\
& 214 (N-H) & 0.0033 & 0.0178 & 0.0048 & 0.0649 \\
  \hline

  MBNNB+NO
& 184 (N-N) & 0.2371 & -0.0092 & 0.7803 & 0.6533 \\
& 199 (B-N) & 0.1337 & 0.3947 & 0.5733 & 0.5176 \\

  MBNNB+NH\textsubscript{3}
& 191 (N-H) & 0.0043 & 0.0235 & 0.0065 & 0.0750 \\
& 214 (N-H) & 0.0042 & 0.0243 & 0.0058 & 0.0710 \\
& 195 (B-H) & 0.0040 & 0.0202 & 0.0065 & 0.0748 \\
& 175 (N-H) & 0.0030 & 0.0159 & 0.0043 & 0.0616 \\

  MBNNB+COCl\textsubscript{2}
& 284 (N-Cl) & 0.0039 & 0.0287 & 0.0032 & 0.0535 \\
& 296 (N-Cl) & 0.0030 & 0.0208 & 0.0026 & 0.0490 \\
& 347 (N-Cl) & 0.0030 & 0.0193 & 0.0028 & 0.0508 \\
& 337 (N-Cl) & 0.0028 & 0.0205 & 0.0022 & 0.0446 \\
& 335 (B-Cl) & 0.0027 & 0.0168 & 0.0025 & 0.0482
\\
& 301 (H-C) & 0.0027 & 0.0145 & 0.0036 & 0.0572 \\
& 341 (B-O) & 0.0020 & 0.0141 & 0.0015 & 0.0369 \\
& 323 (N-C) & 0.0019 & 0.0132 & 0.0014 & 0.0366 \\
& 339 (N-O) & 0.0014 & 0.0105 & 0.0009 & 0.0291  \\
   \hline
\end{tabular}
\begin{flushleft}
\tiny {$^\dagger$ $\rho$, $\nabla ^2\rho$, ELF and LOL are in atomic units.}
\end{flushleft}
\end{center}
\end{table}

\subsection{Molecular dynamics simulation}
\label{Sec:MD}
Molecular dynamics simulations were carried out to elucidate whether the interaction between nanobelts and greenhouse gases was stable over time. In case BNNB interacting with single-molecule gas systems, the complexes BNNB+H\textsubscript{2}S, BNNB+CO and BNNB+CH\textsubscript{4} broke apart, the gas molecule flying off the
nanobelt. For CO and CH\textsubscript{4} it is reasonable because they showed low adsorption energies and small recovery times (see Table~\ref{Tab:ResultsAdsBNNB}). The adsorption energy of CO\textsubscript{2} is lower than for H\textsubscript{2}S but very similar. Therefore, it could be expected that, after the $E_{ads}$ ranking, the CO\textsubscript{2} detached H\textsubscript{2}S. This behavior could be associated with the difference in mass of these molecules, having H\textsubscript{2}S a mass of
34.08~gmol\textsuperscript{-1}, whereas the mass of CO\textsubscript{2} is higher and equal to 44.01~gmol\textsuperscript{-1}. For MBNNB, the complexes
MBNNB+H\textsubscript{2}S, MBNNB+CO and MBNNB+CH\textsubscript{4} also broke apart following the lowest $E_{ads}$ ranking (see Table~\ref{Tab:ResultsAdsMBNNB}). The molecular dynamics animations for BNNB that interacts with single-molecule gas systems can be seen by pointing the cell phone camera to the upper QR code in Figure~\ref{Fig:RDFPackages}.

\renewcommand{\sizeA}{3.0cm}
\newcommand{\sizeB}{11.0cm}
\begin{figure}[tbph]
\centering
\begin{tabular}{cc}
\multirow{2}{*}{\subfigure{\includegraphics[width=\sizeB]{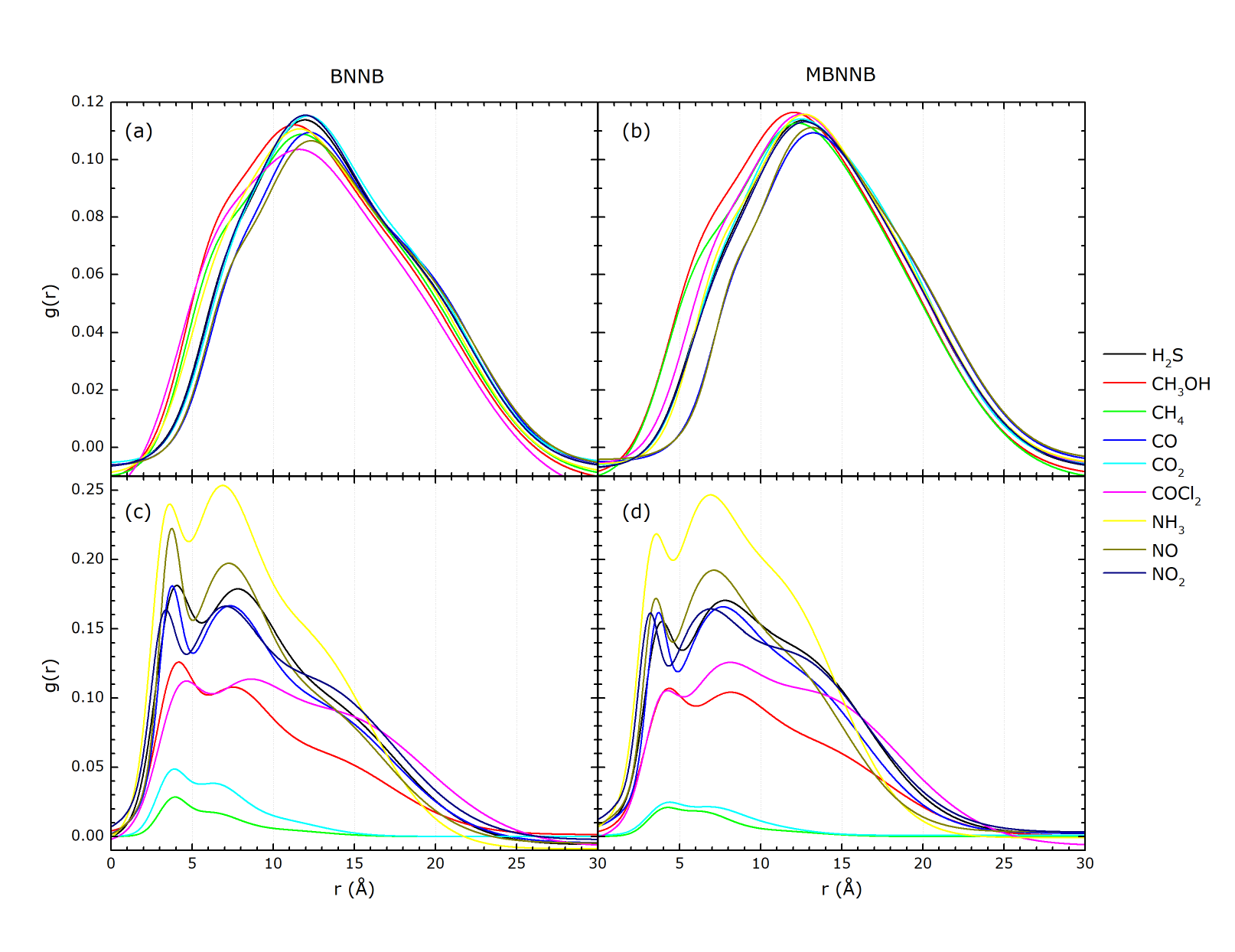}}
} \\ & \subfigure{\includegraphics[width=\sizeA]{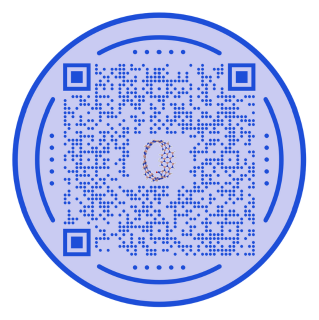}} \\
 & \subfigure{\includegraphics[width=\sizeA]{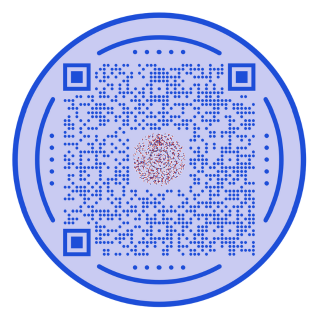}} \\
\end{tabular}
\vspace{1.5cm}
\caption{Initial (a,b) and final (c,d) radial distribution functions for all systems calculated with VMD~\cite{vmd} and TRAVIS~\cite{travis_0,travis_1}. Top QR Code: molecular dynamics animations for nanobelts interacting with single gas molecules. Bottom QR Code: molecular dynamics animations for nanobelts interacting with 500 gas molecules.}
\label{Fig:RDFPackages}
\end{figure}

Figure~\ref{Fig:RDFPackages} shows the calculated radial distribution function (RDF) for packed systems, that is, nanobelts that interact with 500 gas molecules. Panels~\ref{Fig:RDFPackages}(a), and~\ref{Fig:RDFPackages}(b) correspond to the initial complexes created by the PACKMOL software (initial frame). As molecules are randomly added~\cite{packmol_0,packmol_1}, the RDF shows a Gaussian distribution shape with three visible peaks. Panels~\ref{Fig:RDFPackages}(c), and~\ref{Fig:RDFPackages}(d) showed the RDF after 100~ ps simulation time. In this case, the three peaks are also visible. All RDF curves were fitted with three Gaussians, and their peak positions ($r_1$, $r_2$, and $r_3$) are shown in tables~\ref{Tab:ResultsMDBNNB}, and~\ref{Tab:ResultsMDMBNNB} for BNNB and MBNNB, respectively. These peaks represent the shell radii where the particles are found with higher probabilities. In both tables, the top line corresponds to the initial frame ($t=0$~ps) and the bottom line corresponds to the results of the final frame ($t=100$~ ps).

All the shell radii showed a decrease in value. This means that at first sight, both nanobelts behave as attractors for all of the gases studied. To elucidate this, we determine the number of molecules within a radius of 10~\AA, before ($n_b$) and after ($n_a$) the simulation, and then the variation $\Delta n=n_a - n_b$ is calculated. Positive values of $\Delta n$ indicate that nanobelts act as attractors for gas molecules, whereas negative values indicate that the repulsion forces between the gas molecules are greater than the attraction with the nanobelts. Both CH\textsubscript{4} and CO\textsubscript{2} have shown negative values for $\Delta n$, while all other gases have shown positive values. It should be noted that the volume of molecules has an important effect here, as molecules such as COCl\textsubscript{2} and CH\textsubscript{3}OH, well ranked according to the adsorption energy (see Tables~\ref{Tab:ResultsAdsBNNB} and~\ref{Tab:ResultsAdsMBNNB}) present lower positive values $\Delta n$. Finally, the calculation of the PMF shows that the change in free energy is always favorable, as it is always negative for all the complexes. The molecular dynamics animations for MBNNB that interacts with the gas of packed molecules can be seen by pointing the cell phone camera to the lower QR
code in Figure~\ref{Fig:RDFPackages}.

\begin{table}[htpb]
\caption{Shells radii, number of molecules inside a shell with 10~\AA~radius ($n$), percentage of variation on the number of molecules ($\% \Delta n$) in a shell with 10~\AA~radius after 100~ps MD simulation time, and PMF ($W$) at first shell for BNNB complexes$^\dagger$.}
\label{Tab:ResultsMDBNNB}
\begin{center}
\renewcommand{\arraystretch}{0.7}
\begin{tabular}{lrrrrrr}
  \hline
  System & $r_1$ & $r_2$ & $r_3$ & $n$ & $\Delta n$ & $W(r_1)$ \\
\hline
  \hline
  BNNB+NO & 6.938 & 11.487 & 18.728 & 33 & --- & --- \\
  & 3.613 & 6.712 & 12.163 & 90 & 173 & -5.679 \\
\hline
  BNNB+NH\textsubscript{3}& 6.208 & 10.659 & 17.457 & 42 & --- & --- \\
  & 3.278 & 6.253 & 11.530 & 115 & 174 & -5.972 \\
\hline
  BNNB+CH\textsubscript{3}OH  & 5.896 & 10.115 & 17.073 & 47 & --- & --- \\
  & 3.830 & 7.004 & 12.502 & 50 & 6 & -3.167 \\
\hline
  BNNB+COCl\textsubscript{2}  & 5.795 & 10.459 & 16.435 & 47 & --- & --- \\
  & 4.051 & 7.437 & 13.862 & 57 & 21 & -2.758 \\
\hline
  BNNB+NO\textsubscript{2}& 6.844 & 11.305 & 18.670 & 38 & --- & --- \\
  & 3.172 & 6.298 & 12.274 & 77 & 103 & -4.139 \\
\hline
  BNNB+H\textsubscript{2}S& 6.717 & 11.225 & 18.650 & 38 & --- & --- \\
  & 3.698 & 7.124 & 12.781 & 85 & 124 & -4.532 \\
\hline
  BNNB+CO\textsubscript{2}& 7.041 & 11.444 & 18.497 & 37 & --- & --- \\
  & 3.636 & 6.183 & 9.688 & 15 & -59 & -1.226 \\
\hline
  BNNB+CO & 7.076 & 11.525 & 18.853 & 34 & --- & --- \\
  & 3.614 & 6.697 & 6.697 & 76 & 124 & -4.610 \\
\hline
  BNNB+CH\textsubscript{4}& 5.968 & 5.968 & 17.051 & 42 & --- & --- \\
  & 3.752 & 6.001 & 9.870 & 7 & -83 & -0.723 \\
\hline
\hline
\end{tabular}
\begin{flushleft}
\tiny {$^\dagger$ Radii are in units of \AA, and $W$ is in units of $eV$.}
\end{flushleft}
\end{center}
\end{table}

\begin{table}[htpb]
\caption{Shells radii, number of molecules inside a shell with 10~\AA~radius ($n$), percentage of variation on the number of molecules ($\% \Delta n$) in a shell with 10~\AA~radius after 100~ps MD simulation time, and PMF ($W$) at first shell for MBNNB complexes$^\dagger$.}
\label{Tab:ResultsMDMBNNB}
\begin{center}
\renewcommand{\arraystretch}{0.7}
\begin{tabular}{lrrrrrr}
  \hline
  System & $r_1$ & $r_2$ & $r_3$ & $n$ & $\Delta n$ & $W(r_1)$ \\
  \hline
  \hline
  MBNNB+NO & 8.018 & 11.855 & 17.663 & 25 & --- & --- \\
   & 3.372 & 6.288 & 10.950 & 86 & 244 & -4.356 \\
  \hline
  MBNNB+NH\textsubscript{3}& 7.051 & 11.287 & 16.866 & 35 & --- & --- \\
   & 3.229 & 5.972 & 10.643 & 114 & 226 & -5.426 \\
  \hline
  MBNNB+COCl\textsubscript{2}  & 6.643 & 11.169 & 17.243 & 39 & --- & --- \\
   & 3.736 & 13.173 & 6.953 & 60 & 54 & -2.580 \\
  \hline
  MBNNB+CH\textsubscript{3}OH  & 5.807 & 10.399 & 15.792 & 45 & --- & --- \\
   & 3.913 & 7.301 & 12.793 & 49 & 9 & -2.664 \\
  \hline
  MBNNB+NO\textsubscript{2}& 6.730 & 11.191 & 16.940 & 35 & --- & --- \\
   & 3.059 & 5.957 & 11.704 & 76 & 117 & -4.109 \\
  \hline
  MBNNB+CO\textsubscript{2}& 6.748 & 10.650 & 15.966 & 35 & ---  & --- \\
   & 3.959 & 6.581 & 9.333 & 9 & -74 & -0.613 \\
  \hline
  MBNNB+CO & 8.008 & 11.800 & 17.519 & 26 & --- & --- \\
   & 3.575 & 6.699 & 11.796 & 75 & 188 & -4.125 \\
  \hline
  MBNNB+H\textsubscript{2}S& 6.670 & 10.940 & 16.406 & 35 & --- & --- \\
   & 3.620 & 11.970 & 6.726 & 78 & 123 & -3.892 \\
  \hline
  MBNNB+CH\textsubscript{4}& 5.612 & 10.533 & 15.836 & 43 & --- & --- \\
   & 3.844 & 6.129 & 9.906 & 7 & -84 & -0.512 \\
  \hline
\end{tabular}
\begin{flushleft}
\tiny {$^\dagger$ Radii are in units of \AA, and $W$ is in units of $eV$.}
\end{flushleft}
\end{center}
\end{table}

\section{Conclusions}
\label{Sec:Conclusions}
We can conclude that the answer to the title question is affirmative based on our findings. We provide evidence that boron-nitride nanobelts exhibit varying degrees of sensitivity and specificity in their response to each of the greenhouse gases on the list, in addition to their capability of capturing them.

The adsorption energy of both nanobelts, determined through optimized geometry calculations, is negative for all gases. However, it is worth noting that the Möbius boron-nitride nanobelt possesses lower adsorption energies for all gaseous substances. The distinct values of the variations in electrical conductivity enable the identification of the interacting gas using nanobelts. It is feasible to reuse belts that have been adsorbed with gases, since the recovery time for the highest recovery time is approximately two hours, whereas the recovery times for the remaining systems are in seconds or even less.

By combining electronic calculations and topological studies, we are able to clarify the manner in which the BNNB and MBNNB belts interact with the enumerated greenhouse gases. Strong covalent bonds can be formed between the nitrogen gas atom and the nitrogen/boron atoms from the belts in the case of NO gas because of the interaction with both nanobelts. By forming a bond between atoms at opposite positions (from the twisted region), this interaction can close the Möbius nanobelt. Topology studies have identified the formation of ring and cage critical points in addition to bond critical points for all of the top-ranked gases, which indicates that the structure of the electronic belts has been altered.  BNNB and MBNNB are subject to noncovalent bonding with the remaining gases.

The stability of each complex was subsequently investigated through the
utilization of molecular dynamic simulations. There were two types of MD performed. The initial type of investigation focused on the stability of the belts when a single gas atom was used, while the subsequent type examined the behavior of nanobelts when in contact with 500 gas molecules. The complexes BNNB+H\textsubscript{2}S, BNNB+CO, BNNB+CH\textsubscript{4}, MBNNB+H\textsubscript{2}S, MBNNB+CO and MBNNB+CH\textsubscript{4} disintegrated upon initial examination, with the corresponding gases escaping from the belts. These gases are classified as gases with lower interactions. The second variant of MD demonstrated a reduction in the radius of the first shell, suggesting a favorable interaction, despite the fact that the number of gas molecules within a sphere of 10~\AA radius may decrease in certain cases. The free energy change is consistently negative, indicating a favorable interaction between the nanobelts and each greenhouse gas.

\section*{CRediT authorship contribution statement}

\textbf{C. Aguiar}: Investigation, Formal analysis, Writing-original draft, Writing-review \& editing.

\textbf{I. Camps}: Conceptualization, Methodology, Software, Formal analysis, Resources, Writing-review \& editing, Supervision, Project administration.

\section*{Declaration of competing interest}

The authors declare that they have no known competing financial interests or personal relationships that could have appeared to influence the work reported in this paper.

\section*{Data availability}
The raw data required to reproduce these findings are available to download from \\
\href{https://doi.org/10.5281/zenodo.10674174}{https://doi.org/10.5281/zenodo.10674174}.

\section*{Funding}
The authors declare that no funds, grants, or other support was received during the preparation of this manuscript.

\section*{Acknowledgements}
We would like to acknowledge financial support from the Brazilian agencies CNPq, CAPES and FAPEMIG. Part of the results presented here were developed with the help of a CENAPAD-SP (Centro Nacional de Processamento de Alto Desempenho em S\~ao Paulo) grant UNICAMP/FINEP-MCT, CENAPAD-UFC (Centro Nacional de Processamento de Alto Desempenho, at Universidade Federal do Cear\'a) and Digital Research Alliance of
Canada (via  project bmh-491-09 belonging to Dr. Nike Dattani), for
computational support.

\newpage

\begin{thebibliography}{10}
\expandafter\ifx\csname url\endcsname\relax
\def\url#1{\texttt{#1}}\fi
\expandafter\ifx\csname urlprefix\endcsname\relax\def\urlprefix{URL }\fi
\expandafter\ifx\csname href\endcsname\relax
\def\href#1#2{#2} \def\path#1{#1}\fi

\bibitem{Lee-Sens.ActuatorsBChem.-255-1788-2018}
S.~W. Lee, W.~Lee, Y.~Hong, G.~Lee, D.~S. Yoon, Recent advances in carbon
material-based {NO\textsubscript{2}} gas sensors, Sens. Actuators B Chem. 255
(2018) 1788--1804 (2018).
\newblock \href {https://doi.org/10.1016/j.snb.2017.08.203}
{\path{doi:10.1016/j.snb.2017.08.203}}.

\bibitem{Manisalidis-Front.PublicHealth-8--2020}
I.~Manisalidis, E.~Stavropoulou, A.~Stavropoulos, E.~Bezirtzoglou,
Environmental and health impacts of air pollution: {A} review, Front. Public
Health 8 (2020).
\newblock \href {https://doi.org/10.3389/fpubh.2020.00014}
{\path{doi:10.3389/fpubh.2020.00014}}.

\bibitem{Nool-SSM-PopulationHealth-15-100879-2021}
C.~No\"ol, C.~Vanroelen, S.~Gadeyne, Qualitative research about public health
risk perceptions on ambient air pollution. a review study, SSM - Population
Health 15 (2021) 100879 (2021).
\newblock \href {https://doi.org/10.1016/j.ssmph.2021.100879}
{\path{doi:10.1016/j.ssmph.2021.100879}}.

\bibitem{Holt-Intern.Med.J.-48-335-2018}
N.~R. Holt, C.~P. Nickson, Severe methanol poisoning with neurological
sequelae: implications for diagnosis and management, Intern. Med. J. 48
(2018) 335--339 (2018).
\newblock \href {https://doi.org/10.1111/imj.13725}
{\path{doi:10.1111/imj.13725}}.

\bibitem{Jo-Tuberc.Respir.Dis.-74-120-2013}
J.~Y. Jo, Y.~S. Kwon, J.~W. Lee, J.~S. Park, B.~H. Rho, W.-I. Choi, Acute
respiratory distress due to methane inhalation, Tuberc. Respir. Dis. 74
(2013) 120 (2013).
\newblock \href {https://doi.org/10.4046/trd.2013.74.3.120}
{\path{doi:10.4046/trd.2013.74.3.120}}.

\bibitem{Buboltz2023}
J.~B. Buboltz, M.~Robins, Hyperbaric treatment of carbon monoxide toxicity
(2023).

\bibitem{Otterness-Emerg.Med.Pract.-20-1-2018}
K.~Otterness, C.~Ahn, Emergency department management of smoke inhalation
injury in adults, Emerg. Med. Pract. 20 (2018) 1--24 (2018).

\bibitem{Schrier-Front.Toxicol.-4--2022}
R.~van~der Schrier, M.~van Velzen, M.~Roozekrans, E.~Sarton, E.~Olofsen,
M.~Niesters, C.~Smulders, A.~Dahan, Carbon dioxide tolerability and toxicity
in rat and man: {A} translational study, Front. Toxicol. 4 (2022).
\newblock \href {https://doi.org/10.3389/ftox.2022.1001709}
{\path{doi:10.3389/ftox.2022.1001709}}.

\bibitem{Rendell-Toxicol.Lett.-290-145-2018}
R.~Rendell, S.~Fairhall, S.~Graham, S.~Rutter, P.~Auton, A.~Smith, R.~Perrott,
B.~Jugg, Assessment of {N}-acetylcysteine as a therapy for phosgene-induced
acute lung injury, Toxicol. Lett. 290 (2018) 145--152 (2018).
\newblock \href {https://doi.org/10.1016/j.toxlet.2018.03.025}
{\path{doi:10.1016/j.toxlet.2018.03.025}}.

\bibitem{Pauluhn-Toxicology-450-152682-2021}
J.~Pauluhn, Phosgene inhalation toxicity: {Update} on mechanisms and
mechanism-based treatment strategies, Toxicology 450 (2021) 152682 (2021).
\newblock \href {https://doi.org/10.1016/j.tox.2021.152682}
{\path{doi:10.1016/j.tox.2021.152682}}.

\bibitem{Ng-J.Med.Toxicol.-15-287-2019}
P.~C. Ng, T.~B. Hendry-Hofer, A.~E. Witeof, M.~Brenner, S.~B. Mahon, G.~R.
Boss, P.~Haouzi, V.~S. Bebarta, Hydrogen sulfide toxicity: {Mechanism} of
action, clinical presentation, and countermeasure development, J. Med.
Toxicol. 15 (2019) 287--294 (2019).
\newblock \href {https://doi.org/10.1007/s13181-019-00710-5}
{\path{doi:10.1007/s13181-019-00710-5}}.

\bibitem{Pangeni-AnnalsofMedicine&Surgery-82-104741-2022}
R.~P. Pangeni, B.~Timilsina, P.~R. Oli, S.~Khadka, P.~R. Regmi, A
multidisciplinary approach to accidental inhalational ammonia injury: A case
report, Annals of Medicine \& Surgery 82 (2022) 104741 (2022).
\newblock \href {https://doi.org/10.1016/j.amsu.2022.104741}
{\path{doi:10.1016/j.amsu.2022.104741}}.

\bibitem{Amaducci2023}
A.~Amaducci, J.~W. Downs, Nitrogen dioxide toxicity (2023).

\bibitem{Verma-ACSSensors-8-3320-2023}
G.~Verma, A.~Gokarna, H.~Kadiri, K.~Nomenyo, G.~Lerondel, A.~Gupta, Multiplexed
gas sensor: {Fabrication} strategies, recent progress, and challenges, ACS
Sensors 8 (2023) 3320--3337 (2023).
\newblock \href {https://doi.org/10.1021/acssensors.3c01244}
{\path{doi:10.1021/acssensors.3c01244}}.

\bibitem{Nazemi-Sensors-19-1285-2019}
H.~Nazemi, A.~Joseph, J.~Park, A.~Emadi, Advanced micro- and nano-gas sensor
technology: {A} review, Sensors 19 (2019) 1285 (2019).
\newblock \href {https://doi.org/10.3390/s19061285}
{\path{doi:10.3390/s19061285}}.

\bibitem{Abooali-J.Comput.Electron.-19-1373-2020}
A.~Abooali, F.~Safari, Adsorption and optical properties of
{H\textsubscript{2}S}, {CH\textsubscript{4}}, {NO}, and {SO\textsubscript{2}}
gas molecules on arsenene: a {DFT} study, J. Comput. Electron. 19 (2020)
1373--1379 (2020).
\newblock \href {https://doi.org/10.1007/s10825-020-01565-8}
{\path{doi:10.1007/s10825-020-01565-8}}.

\bibitem{Ahmed-R.Soc.OpenSci.-9-220778-2022}
M.~T. Ahmed, S.~Islam, F.~Ahmed, Density functional theory study of {Mobius}
boron-carbon-nitride as potential {CH\textsubscript{4}} ,
{H\textsubscript{2}S}, {NH\textsubscript{3}}, {COCl\textsubscript{2}} and
{CH\textsubscript{3}OH} gas sensor, R. Soc. Open Sci. 9 (2022) 220778 (2022).
\newblock \href {https://doi.org/10.1098/rsos.220778}
{\path{doi:10.1098/rsos.220778}}.

\bibitem{Calvaresi-J.Mater.Chem.A-2-12123-2014a}
M.~Calvaresi, F.~Zerbetto, Atomistic molecular dynamics simulations reveal
insights into adsorption, packing, and fluxes of molecules with carbon
nanotubes, J. Mater. Chem. A 2 (2014) 12123--12135 (2014).
\newblock \href {https://doi.org/10.1039/c4ta00662c}
{\path{doi:10.1039/c4ta00662c}}.

\bibitem{Cezar-arXiv-2023}
H.~M. Cezar, T.~D. Lanna, D.~A. Damasceno, A.~Kirch, C.~R. Miranda, Revisiting
greenhouse gases adsorption in carbon nanostructures: advances through a
combined first-principles and molecular simulation approach, arXiv (2023).
\newblock \href {https://doi.org/10.48550/arXiv.2307.11710}
{\path{doi:10.48550/arXiv.2307.11710}}.

\bibitem{Chang-ACSNano-4-5095-2010}
C.-C. Chang, I.-K. Hsu, M.~Aykol, W.-H. Hung, C.-C. Chen, S.~B. Cronin, A new
lower limit for the ultimate breaking strain of carbon nanotubes, ACS Nano 4
(2010) 5095--5100 (2010).
\newblock \href {https://doi.org/10.1021/nn100946q}
{\path{doi:10.1021/nn100946q}}.

\bibitem{Holt-Science-312-1034-2006}
J.~K. Holt, H.~G. Park, Y.~Wang, M.~Stadermann, A.~B. Artyukhin, C.~P.
Grigoropoulos, A.~Noy, O.~Bakajin, Fast mass transport through
sub-2-nanometer carbon nanotubes, Science 312 (2006) 1034--1037 (2006).
\newblock \href {https://doi.org/10.1126/science.1126298}
{\path{doi:10.1126/science.1126298}}.

\bibitem{Safari-Appl.Surf.Sci.-464-153-2019}
F.~Safari, M.~Moradinasab, M.~Fathipour, H.~Kosina, Adsorption of the
{NH\textsubscript{3}}, {NO}, {NO\textsubscript{2}}, {CO\textsubscript{2}},
and {CO} gas molecules on blue phosphorene: {A} first-principles study, Appl.
Surf. Sci. 464 (2019) 153--161 (2019).
\newblock \href {https://doi.org/10.1016/j.apsusc.2018.09.048}
{\path{doi:10.1016/j.apsusc.2018.09.048}}.

\bibitem{Tang-Sensors-21-1443-2021}
X.~Tang, M.~Debliquy, D.~Lahem, Y.~Yan, J.-P. Raskin, A review on
functionalized graphene sensors for detection of ammonia, Sensors 21 (2021)
1443 (2021).
\newblock \href {https://doi.org/10.3390/s21041443}
{\path{doi:10.3390/s21041443}}.

\bibitem{Wu-RSCAdvances-14-1445-2024}
P.~Wu, Z.~Zhao, Z.~Huang, M.~Huang, Toxic gas sensing performance of arsenene
functionalized by single atoms {(Ag, Au): a DFT} study, RSC Advances 14
(2024) 1445--1458 (2024).
\newblock \href {https://doi.org/10.1039/d3ra07816g}
{\path{doi:10.1039/d3ra07816g}}.

\bibitem{Oliveira-Sci.Rep.-12-22393-2022}
R.~B. de~Oliveira, D.~D. Borges, L.~D. Machado, Mechanical and gas adsorption
properties of graphene and graphynes under biaxial strain, Sci. Rep. 12
(2022) 22393 (2022).
\newblock \href {https://doi.org/10.1038/s41598-022-27069-y}
{\path{doi:10.1038/s41598-022-27069-y}}.

\bibitem{Li-Mater.Chem.Phys.-301-127602-2023}
C.~Li, Y.~Chen, Z.~Xu, X.~Yang, Quantum mechanical analysis of adsorption for
{CH\textsubscript{4}} and {CO\textsubscript{2}} onto graphene oxides, Mater.
Chem. Phys. 301 (2023) 127602 (2023).
\newblock \href {https://doi.org/10.1016/j.matchemphys.2023.127602}
{\path{doi:10.1016/j.matchemphys.2023.127602}}.

\bibitem{Poudel-Mater.TodayPhys.-7-7-2018}
Y.~R. Poudel, W.~Li, Synthesis, properties, and applications of carbon
nanotubes filled with foreign materials: {A} review, Mater. Today Phys. 7
(2018) 7--34 (2018).
\newblock \href {https://doi.org/10.1016/j.mtphys.2018.10.002}
{\path{doi:10.1016/j.mtphys.2018.10.002}}.

\bibitem{Alexiadis-Chem.Phys.Lett.-460-512-2008}
A.~Alexiadis, S.~Kassinos, Molecular dynamic simulations of carbon nanotubes in
{CO\textsubscript{2}} atmosphere, Chem. Phys. Lett. 460 (2008) 512--516
(2008).
\newblock \href {https://doi.org/10.1016/j.cplett.2008.06.050}
{\path{doi:10.1016/j.cplett.2008.06.050}}.

\bibitem{Lithoxoos-J.ofSupercriticalFluids-55-510-2010}
G.~P. Lithoxoos, A.~Labropoulos, L.~D. Peristeras, N.~Kanellopoulos, J.~Samios,
I.~G. Economou, Adsorption of {N\textsubscript{2}, CH\textsubscript{4}, CO
and CO\textsubscript{2}} gases in single walled carbon nanotubes: {A}
combined experimental and {Monte Carlo} molecular simulation study, J. of
Supercritical Fluids 55 (2010) 510--523 (2010).
\newblock \href {https://doi.org/10.1016/j.supflu.2010.09.017}
{\path{doi:10.1016/j.supflu.2010.09.017}}.

\bibitem{Dillon-Nature-386-377-1997}
A.~C. Dillon, K.~M. Jones, T.~A. Bekkedahl, C.~H. Kiang, D.~S. Bethune, M.~J.
Heben, Storage of hydrogen in single-walled carbon nanotubes, Nature 386
(1997) 377--379 (1997).
\newblock \href {https://doi.org/10.1038/386377a0}
{\path{doi:10.1038/386377a0}}.

\bibitem{Lyu-Nanomaterials-10-255-2020}
J.~Lyu, V.~Kudiiarov, A.~Lider, An overview of the recent progress in
modifications of carbon nanotubes for hydrogen adsorption, Nanomaterials 10
(2020) 255 (2020).
\newblock \href {https://doi.org/10.3390/nano10020255}
{\path{doi:10.3390/nano10020255}}.

\bibitem{Ganji-Commun.Theor.Phys.-53-987-2010}
M.~D. Ganji, M.~Asghary, A.~A. Najafi, Interaction of methane with
single-walled carbon nanotubes: {R}ole of defects, curvature and nanotubes
type, Commun. Theor. Phys. 53 (2010) 987--993 (2010).
\newblock \href {https://doi.org/10.1088/0253-6102/53/5/37}
{\path{doi:10.1088/0253-6102/53/5/37}}.

\bibitem{Lu-Catalysts-12-561-2022}
C.~Lu, P.~Chen, C.~Li, J.~Wang, Study of intermolecular interaction between
small molecules and carbon nanobelt: {Electrostatic}, exchange, dispersive
and inductive forces, Catalysts 12 (2022) 561 (2022).
\newblock \href {https://doi.org/10.3390/catal12050561}
{\path{doi:10.3390/catal12050561}}.

\bibitem{VNL}
{Virtual NanoLab - Atomistix ToolKit. QuantumWise. v2017.1} (2017).

\bibitem{xTB_2}
S.~Grimme, C.~Bannwarth, P.~Shushkov, A robust and accurate tight-binding
quantum chemical method for structures, vibrational frequencies, and
noncovalent interactions of large molecular systems parametrized for all
spd-block elements ({Z=1--86}), J. Chem. Theory Comput. 13 (2017) 1989--2009
(2017).
\newblock \href {https://doi.org/10.1021/acs.jctc.7b00118}
{\path{doi:10.1021/acs.jctc.7b00118}}.

\bibitem{xTB_GFN0}
P.~Pracht, E.~Caldeweyher, S.~Ehlert, S.~Grimme, A robust non-self-consistent
tight-binding quantum chemistry method for large molecules, ChemRxiv (2019)
chemrxiv.8326202.v1 (2019).
\newblock \href {https://doi.org/10.26434/chemrxiv.8326202.v1}
{\path{doi:10.26434/chemrxiv.8326202.v1}}.

\bibitem{xTB_GFN2}
C.~Bannwarth, S.~Ehlert, S.~Grimme, {GFN2--xTB}--{An} accurate and broadly
parametrized self-consistent tight-binding quantum chemical method with
multipole electrostatics and density-dependent dispersion contributions, J.
Chem. Theory Comput. 15 (2019) 1652--1671 (2019).
\newblock \href {https://doi.org/10.1021/acs.jctc.8b01176}
{\path{doi:10.1021/acs.jctc.8b01176}}.

\bibitem{xTB_1}
C.~Bannwarth, E.~Caldeweyher, S.~Ehlert, A.~Hansen, P.~Pracht, J.~Seibert,
S.~Spicher, S.~Grimme, Extended tight-binding quantum chemistry methods,
WIREs Comput. Mol. Sci. 11 (2020) e1493 (2020).
\newblock \href {https://doi.org/10.1002/wcms.1493}
{\path{doi:10.1002/wcms.1493}}.

\bibitem{xTB-dock}
C.~Plett, S.~Grimme, Automated and efficient generation of general molecular
aggregate structures, Angew. Chem. Int. Ed. 62 (2022).
\newblock \href {https://doi.org/10.1002/anie.202214477}
{\path{doi:10.1002/anie.202214477}}.

\bibitem{xTB-IFF}
S.~Grimme, C.~Bannwarth, E.~Caldeweyher, J.~Pisarek, A.~Hansen, A general
intermolecular force field based on tight-binding quantum chemical
calculations, J. Chem. Phys. 147 (2017) 161708 (2017).
\newblock \href {https://doi.org/10.1063/1.4991798}
{\path{doi:10.1063/1.4991798}}.

\bibitem{AbdalkareemJasim-Inorg.Chem.Commun.-146-110158-2022}
S.~Abdalkareem~Jasim, A.~H. Shather, T.~Alawsi, A.~Alexis Ram\'{\i}rez-Coronel,
A.~B. Mahdi, M.~Normatov, M.~Jade Catalan~Opulencia, F.~Kamali, Adsorption
properties of {B12N12, AlB11N12, and GaB11N12} nanostructure in gas and
solvent phase for phenytoin detecting: {A DFT} study, Inorg. Chem. Commun.
146 (2022) 110158 (2022).
\newblock \href {https://doi.org/10.1016/j.inoche.2022.110158}
{\path{doi:10.1016/j.inoche.2022.110158}}.

\bibitem{Goel-EngineeringReports-5--2023}
N.~Goel, K.~Kunal, A.~Kushwaha, M.~Kumar, Metal oxide semiconductors for gas
sensing, Engineering Reports 5 (2023).
\newblock \href {https://doi.org/10.1002/eng2.12604}
{\path{doi:10.1002/eng2.12604}}.

\bibitem{Kittel1996}
C.~Kittel, Introduction to solid state physics, 7th Edition, John Wiley \&
Sons, Inc., 1996 (1996).

\bibitem{xTB-DIPRO}
J.~T. Kohn, N.~Gildemeister, S.~Grimme, D.~Fazzi, A.~Hansen, Efficient
calculation of electronic coupling integrals with the dimer projection method
via a density matrix tight-binding potential, J. Chem. Phys. 159 (2023)
144106 (2023).
\newblock \href {https://doi.org/10.1063/5.0167484}
{\path{doi:10.1063/5.0167484}}.

\bibitem{multiwfn}
T.~Lu, F.~Chen, Multiwfn: {A} multifunctional wavefunction analyzer, J. Comput.
Chem. 33 (2012) 580--592 (2012).
\newblock \href {https://doi.org/10.1002/jcc.22885}
{\path{doi:10.1002/jcc.22885}}.

\bibitem{packmol_0}
J.~M. Mart{\'{\i}}nez, L.~Mart{\'{\i}}nez, Packing optimization for automated
generation of complex system{\textquotesingle}s initial configurations for
molecular dynamics and docking, J. Comput. Chem. 24 (2003) 819--825 (2003).
\newblock \href {https://doi.org/10.1002/jcc.10216}
{\path{doi:10.1002/jcc.10216}}.

\bibitem{packmol_1}
L.~Mart{\'{\i}}nez, R.~Andrade, E.~G. Birgin, J.~M. Mart{\'{\i}}nez, {PACKMOL}:
A package for building initial configurations for molecular dynamics
simulations, J. Comput. Chem. 30 (2009) 2157--2164 (2009).
\newblock \href {https://doi.org/10.1002/jcc.21224}
{\path{doi:10.1002/jcc.21224}}.

\bibitem{jmol}
{Jmol: An open-source Java viewer for chemical structures in {3D}.
http://www.jmol.org/}.

\bibitem{Hansen2006}
J.~P. Hansen, I.~R. McDonald, Theory of simple liquids, 3rd Edition, Elsevier
Science \& Technology, London, 2006, description based on publisher supplied
metadata and other sources. (2006).

\bibitem{Bader1994}
R.~F.~W. Bader, Atoms in molecules: a quantum theory, International series of
monographs on chemistry, Clarendon Press, Oxford, 1994 (1994).

\bibitem{myMethods}
I.~Camps, Methods used in nanostructure modeling (2023).
\newblock \href {https://doi.org/10.48550/arXiv.2303.01226}
{\path{doi:10.48550/arXiv.2303.01226}}.

\bibitem{elf}
A.~D. Becke, K.~E. Edgecombe, A simple measure of electron localization in
atomic and molecular systems, J. Chem. Phys. 92 (1990) 5397--5403 (1990).
\newblock \href {https://doi.org/10.1063/1.458517}
{\path{doi:10.1063/1.458517}}.

\bibitem{elf2}
K.~Koumpouras, J.~A. Larsson, Distinguishing between chemical bonding and
physical binding using electron localization function ({ELF}), J. Phys.:
Condens. Matter 32 (2020) 315502 (2020).
\newblock \href {https://doi.org/10.1088/1361-648X/ab7fd8}
{\path{doi:10.1088/1361-648X/ab7fd8}}.

\bibitem{lol}
H.~L. Schmider, A.~D. Becke, Chemical content of the kinetic energy density, J.
Mol. Struct.: {THEOCHEM} 527 (2000) 51--61 (2000).
\newblock \href {https://doi.org/10.1016/S0166-1280(00)00477-2}
{\path{doi:10.1016/S0166-1280(00)00477-2}}.

\bibitem{vmd}
W.~Humphrey, A.~Dalke, K.~Schulten, {VMD}: Visual molecular dynamics, Journal
of Molecular Graphics 14 (1996) 33--38 (1996).
\newblock \href {https://doi.org/10.1016/0263-7855(96)00018-5}
{\path{doi:10.1016/0263-7855(96)00018-5}}.

\bibitem{Aguiar-Phys.BCondens.Matter-668-415178-2023}
C.~Aguiar, N.~Dattani, I.~Camps, M\"obius boron-nitride nanobelts interacting
with heavy metal nanoclusters, Phys. B Condens. Matter 668 (2023) 415178
(2023).
\newblock \href {https://doi.org/10.1016/j.physb.2023.415178}
{\path{doi:10.1016/j.physb.2023.415178}}.

\bibitem{Aguiar-J.Mol.Model.-29-277-2023}
C.~Aguiar, N.~Dattani, I.~Camps, M\"obius carbon nanobelts interacting with
heavy metal nanoclusters, J. Mol. Model. 29 (2023) 277 (2023).
\newblock \href {https://doi.org/10.1007/s00894-023-05669-3}
{\path{doi:10.1007/s00894-023-05669-3}}.

\bibitem{Jeffrey1997}
G.~A. Jeffrey, An introduction to hydrogen bonding, Topics in Physical
Chemistry, Oxford University Press, 1997 (1997).

\bibitem{Desiraju1999}
G.~R. Desiraju, T.~Steiner, The weak hydrogen bond in structural chemistry and
biology, Oxford Science Publications, 1999 (1999).

\bibitem{travis_0}
M.~Brehm, B.~Kirchner, {TRAVIS - A} free analyzer and visualizer for {Monte
Carlo} and molecular dynamics trajectories, J. Chem. Inf. Model. 51 (2011)
2007--2023 (2011).
\newblock \href {https://doi.org/10.1021/ci200217w}
{\path{doi:10.1021/ci200217w}}.

\bibitem{travis_1}
M.~Brehm, M.~Thomas, S.~Gehrke, B.~Kirchner, {TRAVIS-A} free analyzer for
trajectories from molecular simulation, J. Chem. Phys. 152 (2020).
\newblock \href {https://doi.org/10.1063/5.0005078}
{\path{doi:10.1063/5.0005078}}.

\end{thebibliography}

\end{document}